\documentclass[prb,preprint]{revtex4}
\usepackage[latin1]{inputenc}
\usepackage{graphicx}
\usepackage{amssymb}
\usepackage{amsmath}
\usepackage{xspace}
\usepackage{dcolumn}
\usepackage{bm}
\usepackage{times}

\begin{document}

\title[Correlation effects in electron-hole plasmas]{
Correlation effects in partially ionized
mass asymmetric electron-hole plasmas}

\author{V.S.~Filinov$^{\dag}$,
H.~Fehske$^\P$, M.~Bonitz$^\ddag$,
V.E.~Fortov$^\dag$, P.~Levashov$^\dag$}
\address{\dag\
Institute for High Energy Density, Russian Academy of Sciences,
Izhorskaya 13/19, Moscow 127412, Russia
}
\address{\P\
Institut f\"ur Physik, Ernst-Moritz-Arndt-Universit{\"a}t
Greifswald, Domstrasse 10a, D-17489 Greifswald, Germany }
\address{\ddag\
Christian-Albrechts-Universit{\"a}t zu Kiel, Institut f\"ur
Theoretische Physik und Astrophysik, Leibnizstrasse 15, 24098 Kiel, Germany
}
\begin{abstract}
The effects of strong Coulomb correlations in dense 
three-dimensional electron-hole plasmas are studied 
by means of unbiased direct path integral Monte Carlo simulations.
The formation and dissociation of bound states, such as
excitons and bi-excitons is analyzed and the density-temperature 
region of their appearance is identified. 
At high density, the Mott transition to the fully ionized
metallic state (electron-hole liquid) is detected. 
Particular attention is paid to the
influence of the hole to electron mass ratio $M$ on 
the properties of the plasma.
Above a critical value of about $M=80$ 
formation of a hole Coulomb crystal was recently
verified [Phys. Rev. Lett. {\bf 95}, 235006 (2005)] 
which is supported by additional results. Results are related to the
excitonic phase diagram of intermediate valent Tm[Se,Te], 
where large values of $M$ have been observed experimentally.
\end{abstract}

\maketitle
\section{Introduction}
Strongly correlated Coulomb systems have been in the focus of
recent investigations in many fields, including dense plasmas in space and
laboratory \cite{boston97,binz96,green-book,ktnp04}, electron-hole plasmas
in semiconductors, charged particles confined in traps or storage
rings (see e.g.~\cite{dubin99} for an overview). In these systems the Coulomb
interaction energy $U$ is often larger than the mean kinetic energy $K$, 
i.e. the coupling parameter 
$\Gamma=|\langle U \rangle|/\langle K \rangle > 1$. 
recently  Coulomb and Wigner crystallization, 
which may occur when $\Gamma\sim 100 \gg 1$,  
has attracted much attention. 
Coulomb crystals were observed in ultracold trapped 
ions~\cite{WBI87,DPC87,itano98}, in dusty plasmas 
and storage rings~\cite{HaTa94,ThMo96,bonitz-etal.06prl}.

There exist many strongly correlated Coulomb systems where 
quantum effects are important. Examples are dense astrophysical plasmas in the
interior of giant planets or white dwarf stars \cite{segretain} as well as
electron-hole plasmas in solids, few-particle electron or exciton clusters in mesoscopic quantum dots (see~\cite{afilinov-etal.01prl,afilinov-etal.03jpa} and references therein).  The formation of Coulomb bound states 
such as atoms and molecules or excitons and bi-excitons, 
of Coulomb liquids and electron-hole droplets \cite{keldysh} exemplify 
the large variety of correlation phenomena that exist in these systems.

Recombination of electrons and positive charges to neutral bound complexes
strongly reduces the Coulomb coupling and thus acts against formation of
Coulomb crystals in two-component charged particle systems.
Quite recently the conditions for the existence of Coulomb crystals in
neutral plasmas containing (at least) two oppositely charged
components has been studied~\cite{Bonitz_PRL05,bonitz-etal.06jpa}, and it was
found that the mass ratio $M$ of positive and negative carriers plays a 
crucial role.
There exists a threshold value of $M$ of about 80 in three-dimensional plasmas.
Such values are possible in semiconductors, which leads to the prediction of
hole crystallization in semiconducting materials with a sufficiently large effective mass
asymmetry~\cite{rice68,mcr_data,Bonitz_PRL05,bonitz-etal.06jpa}. Such values are feasible e.g. in the intermediate valence  Tm[Se,Te] system, 
which under pressure and at very low temperature even might show 
the phenomenon of excitonic Bose condensation~\cite{wachter}.

In the present paper we substantially extend the analysis 
of previous work~\cite{Bonitz_PRL05,bonitz-etal.06jpa} 
on two-component partially ionized Coulomb systems with mass ratios 
$M$ varying between one and about one thousand, thus 
covering plasmas, ranging from positronium, over condensed matter 
systems (almost) to hydrogen.
Thereby we focus on the fundamental aspects of Coulomb 
correlations in two-component plasmas in dependence on $M$. 
Special emphasis is placed on situations with $M$ close to 
the critical value for hole crystallization.

From a theoretical point of view, the complex 
processes of interest which involve
strong Coulomb forces, quantum and spin effects 
are difficult to treat within the framework of analytical
approaches. 
Therefore, over the last decade there has been a high activity in the 
development of numerical techniques capable to tackle 
strongly correlated Coulomb systems (plasmas)~\cite{egger99,KTR94,EbSc97,bonitz-etal.96jpcm,bonitz-book}.
A technique which is particularly well suited to describe
equilibrium properties of two-component plasmas in the strong
coupling and degeneracy regime, is the path integral quantum Monte
Carlo (PIMC) method. Remarkable progress has been obtained in
applying these techniques to 
Fermi systems~\cite{boston97,binz96,zamalin,binder96,berne98}. Since
PIMC simulations of macroscopic Coulomb systems are hampered
by the notorious fermion sign problem, several strategies have been
developed to overcome or at least ``weaken'' this 
difficulty~\cite{egger99,ceperley95,ceperley95rmp}. Within the
restricted PIMC approach additional assumptions on the
density operator were adopted, which reduce the sum over
permutations to even (positive) contributions 
only~\cite{ceperley95,ceperley95rmp}. This requires
the knowledge of the nodes of the density matrix, however, which for
interacting macroscopic systems are known only
approximately~\cite{mil-pol}. 
Hence the accuracy of the results, in particular 
in the regime of strong correlations, is difficult
to assess. An alternative approach are direct PIMC simulations
which have occasionally been attempted by various
groups, e.g.~\cite{imada84} but in general were not sufficiently precise
and efficient for practical purposes. In recent years an improved 
path integral representation of the $N$-particle density operator has been
developed~\cite{filinov-etal.00jetpl,filinov-etal.00pla,FiBoEbFo01} that 
allows for {\it direct fermionic path integral Monte Carlo (DPIMC)}
simulations of dense plasmas in a large range of temperatures
and densities, see \cite{numbook} for an introduction.

The present paper applies the DPIMC method to electron-hole plasmas
with strong mass-asymmetry. Here we consider situations where also 
the heavy component (referred to as ``holes'' hereafter) 
has to be treated quantum mechanically, unlike for hydrogen-like plasmas.
A second extension of our previous simulations is an improved treatment
of the exchange-effects which allows us to reach 
higher densities and lower temperatures, as required to study 
the Mott effect and hole crystallization.
Sec.~\ref{theory} gives a brief overview on the DPIMC
approach for calculating thermodynamic quantities.
Details on the derivations of the basic formulas, 
including equation of state and 
energy and a discussion of the quantum pair potentials used in the simulations 
are given in Appendices~\ref{theory_a} and~\ref{theory_b}. 
An extensive numerical study of strongly correlated two-component Coulomb
systems is presented in Sec.~\ref{simulations}. Here we first give 
an overview on possible correlation effects in the 
limits of small and very large mass ratios. Then we consider 
more in detail the semiconductor system which is closest to the 
critical value of $M=80$ for hole crystallization.
In particular, energy and pressure, the microscopic 
electron-hole configurations, the 
fraction of bound states, as well as 
various (spin-dependent) pair distribution functions and
charge structure factors were calculated in a wide density  and temperature
range. Finally, numerical results for the hole crystal are presented.
The main results are summarized in Sec.~\ref{summary}.

\section{Path integral Monte Carlo procedure}\label{theory}
We consider a neutral two-component plasma consisting of $N_e=N_h=N$ electrons and holes in equilibrium with the Hamiltonian,
${\hat H}={\hat K}+{\hat U}^c$,
containing kinetic energy ${\hat K}$ and Coulomb interaction energy
${\hat U}^c$ parts. The thermodynamic properties at
given temperature $T$ and volume $V$ are then 
completely described by the canonical density operator, 
${\hat \rho} = e^{-\beta {\hat H}}/Z$, with the partition function 
\begin{equation}\label{q-def}
Z(N_e,N_h,V;\beta) = \frac{1}{N_e!N_h!} \sum_{\sigma}\int\limits_V
dq \,\rho(q, \sigma ;\beta),
\end{equation}
where $\beta=1/k_BT$, and $\rho(q, \sigma ;\beta)$ denotes the diagonal matrix
elements of the density operator at a given value $\sigma$ of  total spin
$z-$projection. In Eq.~(\ref{q-def}), 
$q=\{q_e,q_h\}$ and $\sigma=\{\sigma_e,\sigma_h\}$
are the spatial coordinates and spin degrees of freedom
of the electrons and holes, i.e.
$q_a=\{q_{1,a}\ldots q_{l,a}\ldots q_{N_a,a}\}$
and $\sigma_a=\{\sigma_{1,a}\ldots \sigma_{l,a}\ldots
\sigma_{N_a,a}\}$, with $a=e,p$.
All thermodynamic functions can be directly computed from the partition function. The 
resulting equations for density matrix, pressure (equation of state) 
and internal energy will be derived in 
Appendices~\ref{theory_a} and~\ref{theory_b}.

Of course, the exact density matrix of interacting quantum
systems is not known (particularly for low temperatures and high
densities), but it can be constructed using a path integral
approach~\cite{feynman-hibbs} based on the operator identity
$e^{-\beta {\hat H}}= e^{-\Delta \beta {\hat H}}\cdot
e^{-\Delta \beta {\hat H}}\dots  e^{-\Delta \beta {\hat H}}$,
where $\Delta \beta = \beta/(n+1)$. This allows us to express the 
density operator in terms of a product of $(n+1)$ 
known high-temperature density operators 
(at $(n+1)$--times higher temperature). In the coordinate 
representation this yields 
products of off-diagonal high-temperature density matrices 
$\langle q^{(k-1)}|e^{-\Delta \beta {\hat H}}|q^{(k)}\rangle$ 
where $k=1,\dots ,(n+1)$.
Accordingly each particle is represented by a set of $(n+1)$ 
coordinates (``beads''), i.e. the
whole configuration of the particles is represented by a
$3(N_e+N_h)(n+1)$-dimensional vector
$\tilde{q}\equiv\{q_{1,e}^{(0)}, \dots q_{1,e}^{(n+1)},
q_{2,e}^{(0)}\ldots q_{2,e}^{(n+1)}, \ldots q_{N_e,e}^{(n+1)};
q_{1,h}^{(0)}\ldots q_{N_h,h}^{(n+1)} \}$.

\begin{figure}[htb]
\begin{center}
\includegraphics[width=5cm,clip=true]{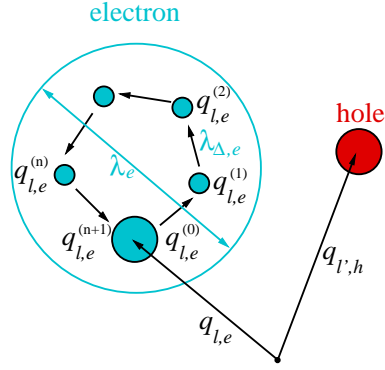}\\
\caption{(Color online) Beads representation of
electrons and holes. Here $\lambda_e^2=2\pi\hbar^2\beta/m_e$,
$\lambda_{\Delta,e}^2=2\pi\hbar^2\Delta\beta/m_e$,
$q^{(1)}_{l,e}=q^{(0)}_{l,e}+\lambda_{\Delta,e}\,\eta^{(1)}_{l,e}$,
and $\sigma=\sigma^{\prime}$. The holes have a similar beads
representation, however $\lambda_{\Delta,h}^2$ is
$(m_h/m_e)$--times smaller, so their beads distribution is not
resolved in the figure.} \label{beads}
\end{center}
\end{figure}
Figure.~\ref{beads} illustrates the representation of one (light)
electron and one (heavy) hole. The circle around the electron
beads symbolizes the region that mainly contributes to the
partition function path integral. The size of this region is of
the order of the thermal electron wavelength $\lambda_e(T)$, while
typical distances  between electron beads are of the order of the
electron wavelength taken at an $(n+1)$--times higher temperature.
The same representation is valid for each hole but it is not shown 
since, due to the larger hole mass, the characteristic length scales
are substantially smaller. Nevertheless, in the simulations below,  
the holes are treated according to the full beads representation. Details, 
including the treatment of the spin, are given in the Appendices. 

To evaluate the density matrix, accurate results for the high-temperature 
approximation are necessary. As we have shown earlier~\cite{FiBoEbFo01}, 
for sufficiently high temperature, i.e. for large number $n$ of time slices) 
each high-temperature factor can be expressed in terms of two-particle
density matrices ($p=1,\dots, N_a, t=1,\dots, N_b, a,b=e,h$)
\begin{eqnarray}\label{rho_ab}
\rho_{ab}(q_{p,a},q'_{p,a}, q_{t,b}, q'_{t,b};\beta)
&=& \frac{(m_a m_b)^{3/2}}{(2 \pi \hbar \beta)^3}
\exp\left[-\frac{m_a}{2 \hbar^2 \beta} (q_{p,a} - q'_{p,a})^2\right]
\, \nonumber\\
&&\times \exp\left[-\frac{m_b}{2 \hbar^2 \beta} (q_{t,b} - q'_{t,b})^2\right]
\exp[-\beta \Phi_{ab}]
\end{eqnarray}
with the familiar Kelbg potential~\cite{Ke63,kelbg}
\begin{eqnarray}
\Phi_{ab}(x_{ab};\beta) =
\frac{e_a e_b}{\lambda_{ab} x_{ab}} \,\left[1-e^{-x_{ab}^2} +
\sqrt{\pi} x_{ab} \left(1-{\rm erf}(x_{ab})\right) \right],
\label{kelbg-d}
\end{eqnarray}
where $x_{ab}=|q_{p,a}-q_{t,b}|/\lambda_{ab}$, and 
the error function ${\rm erf}(x)=\frac{2}{\sqrt{\pi}}\int_0^x
dt e^{-t^2}$. The derivation of approximation 
(\ref{rho_ab}) and discussion of its 
accuracy is given in App.~\ref{theory_b}.

\begin{figure}[htb]
\begin{center}
\includegraphics[width=5cm,clip=true]{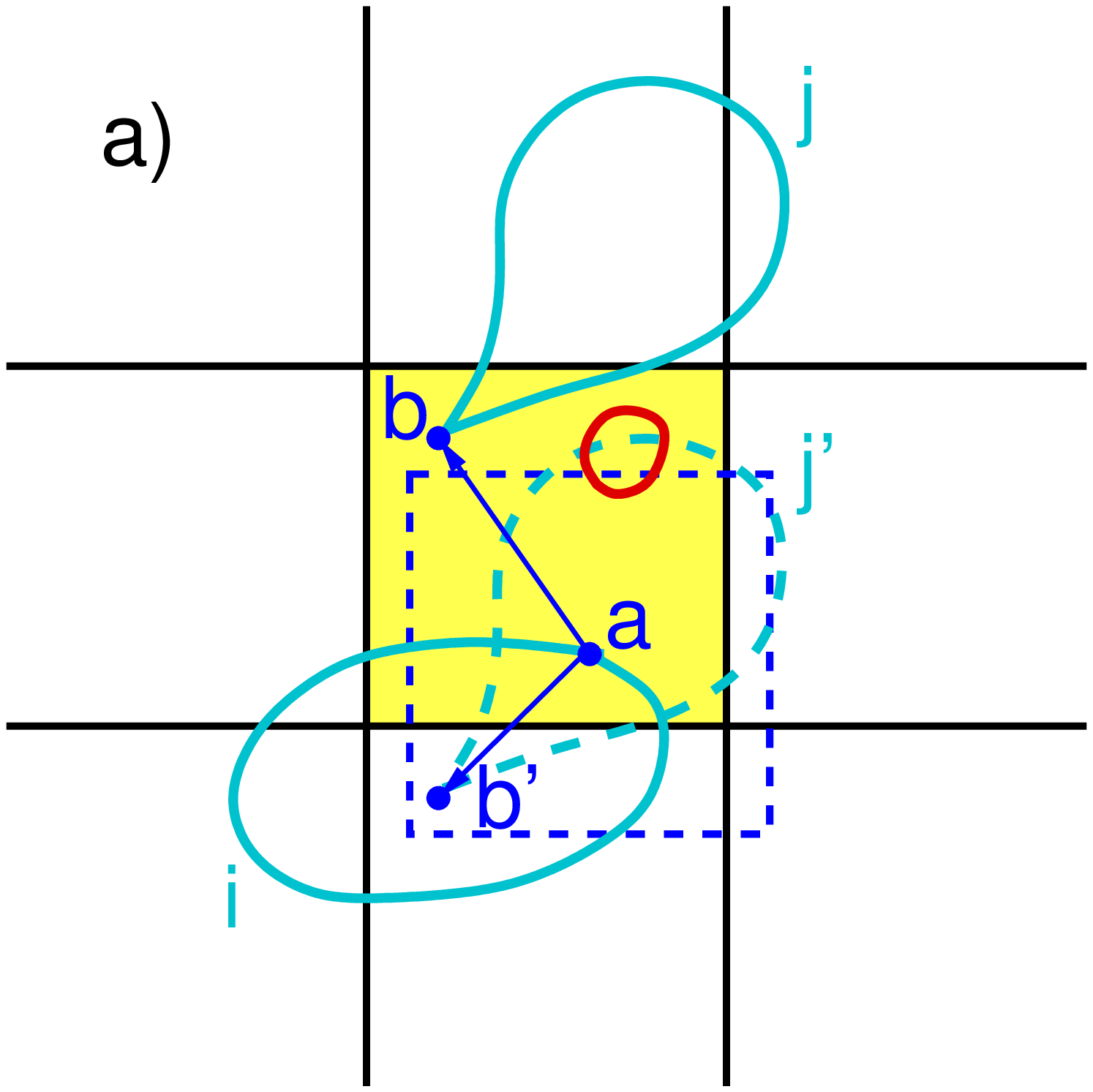}\hspace*{0.3cm}
\includegraphics[width=5cm,clip=true]{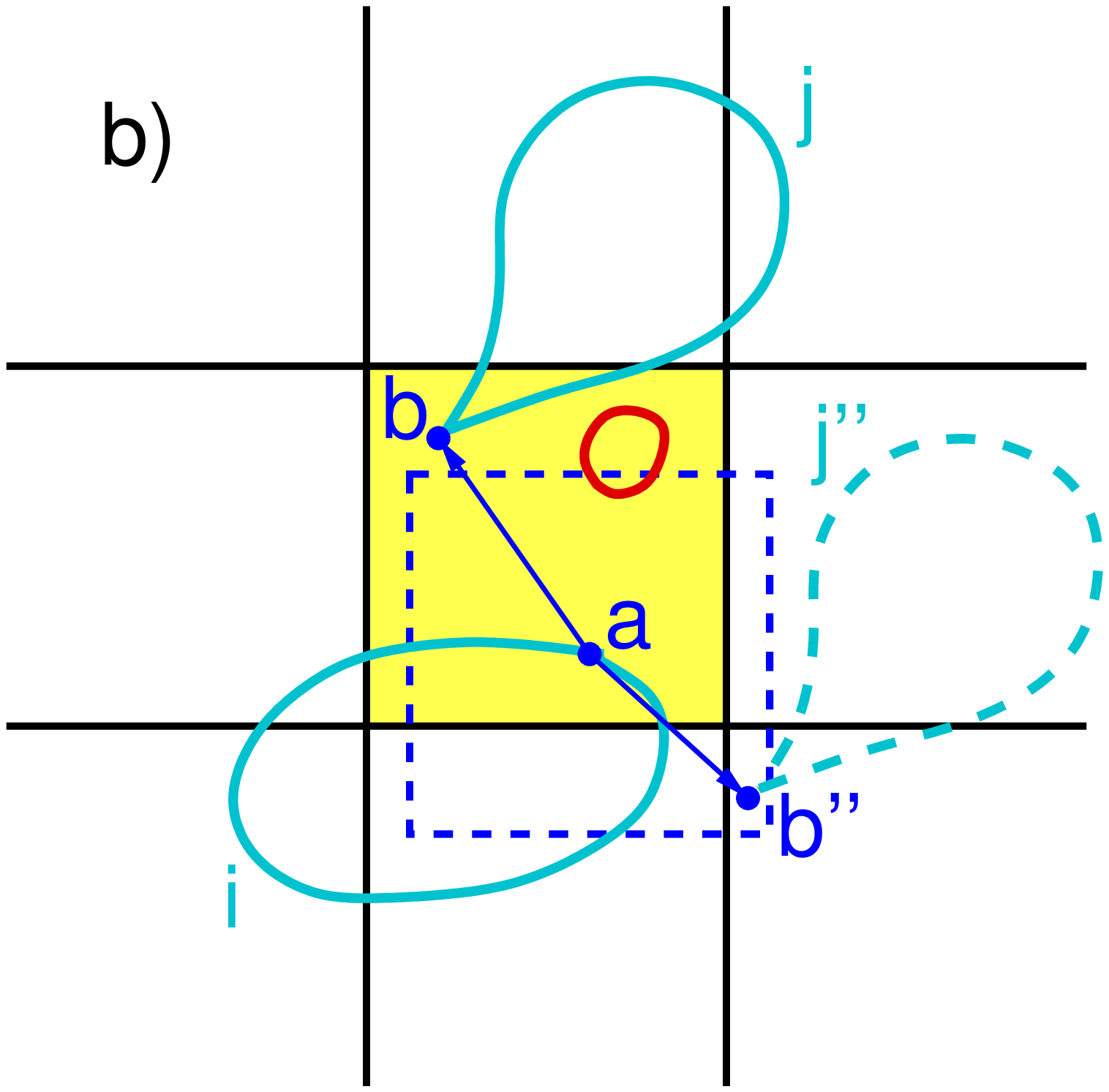}
\caption{(Color online) Sketch of the boundary
conditions used for the simulations.
Electron and hole clouds of beads are denoted
by solid green and red lines respectively. The dashed line
shows the cell for choosing the image of particle $b$ which is closest
to particle $a$. For further explanation see text.}
\label{period}
\end{center}
\end{figure}

In our DPIMC scheme we use different types of steps, 
where either electron (hole) coordinates $q_{t,e}$ ($q_{p,h}$)
or individual electronic (hole) beads are moved until
convergence of the calculated values is reached. Using periodic
boundary conditions (PBC) the basic MC cell (filled yellow square in
Fig.~\ref{period}) is periodically repeated in $x$, $y$ and $z$
directions.

As mentioned above the main contribution to the path integral
representation of the partition function comes from configurations for
which the typical size of the clouds of electronic (hole) beads is of
the order of the thermal wavelength of electrons (holes).
At the moment our computer resources allow to consider 
up to about one hundred electrons and holes with several tens
beads in the basic MC cell. Due to this limitation on the number
of particles in the MC cell there is a restriction on the size of the
MC cell  for a given density.
In the case of a highly degenerate plasma the thermal wavelength and the
typical size of the electronic clouds of beads may be larger than
the size of the basic MC cell. So beads of electrons belonging to
the basic MC cell can penetrate into neighboring images of the main cell,
as well as electronic beads from a neighboring cell can
extend into the basic MC cell. Due to this fact we have improved 
our PIMC scheme compared to previous simulations: In our new
calculations we assume that
an electron (hole) belongs to a certain cell if its physical
coordinate $q^0_{t,a} (t=1,\dots,N_a, a=e,h)$ belongs to this cell.
In Fig.~\ref{period} electron and hole clouds of beads are marked 
by solid green and red lines. Only a few periodic images of these particles
are shown by related dashed lines.

Let us consider the PBC for the calculation of pressure and energy
in more detail. For distances between beads with the same number
$l$ in the Kelbg potentials and its derivatives
[respectively first and second line in curly brackets in Eqs.~(\ref{energy}) and~(\ref{eos})] we used the standard PBC (see Fig.~\ref{period}
a). Namely, in the Kelbg potential and its derivatives, instead of
the distance $\overline{ab}$ between beads with the same 
number $l$ of electrons
$i$ and $j$ we take the smallest distance $\overline{ab'}$  to one of the
electron images $j'$. The same applies to electron-hole and
hole-hole distances. Furthermore, in calculations of the scalar products and
derivatives of the Kelbg potential [terms $C$ and $D$ in Eq.~(\ref{energy})
and terms $A$ and $B$ in Eq.~(\ref{eos})] the situation is more
complicated due to the dependence of the scalar products on the
angle between vectors to beads of the particle from the  basic MC
cell and its periodic images. In our calculations we first of all
choose, for a given particle $i$, the nearest image $j''$, according to
the distance between coordinates $q^0_{i,e}$ and
$q^0_{j'',e}$ only, as shown in Fig.~\ref{period} b). $\overline{ab''}$ is the
smallest distance for all $ab$. Then for this pair $i,j''$ we
calculate all scalar product terms $A,B,C,D$ and 
the related derivatives of the Kelbg
potential. The same is done for electron-hole and hole-hole
pairs.

In our previous calculations
determinants of the exchange matrices [cf. Eq.~(\ref{psi}) 
of App.~\ref{theory_b}]
were only computed for particles belonging to the basic Monte
Carlo cell. However, with increasing degeneracy $(n\lambda^3)$ the
ratio of the  particle thermal wavelength to the size of the Monte
Carlo cell also increases. If this ratio approaches one
exchange effects between particles in the main MC cell and their
images in the neighbor cells have to be included. Therefore, in the present
calculations we take into account the exchange interactions of
electrons and holes from neighboring Monte Carlo cells, namely first
from the  ($3^3-1$) nearest-neighbor cells, then from the $5^3-1$
next nearest-neighbor cells and so on.
These improved calculations were tested
for both an ideal plasma and  a non-ideal hydrogen plasma. Excellent 
agreement with the known analytical results for an ideal plasma
was found up to densities where the
parameter $n\lambda^3$ reaches values of several hundreds.

In the present simulations of dense electron-hole plasmas, we
varied both the particle number and the number of beads. We
found that in order to obtain convergent results for the
thermodynamic properties in the density-temperature range considered below 
it is sufficient to simulate systems with
particle numbers of $N_e=N_h=50 \dots 100$. Of course, the accuracy is
strongly affected by the number of beads $n$. To exclude an 
$n$-dependence of our calculations, the density matrices in the
high-temperature decomposition were always taken at temperatures
above the exciton binding energy. In practice, a number of about
$n=20$ beads turn out to be sufficient. In order to simplify
the computations further, we included only the dominant contribution in
the sums over the total electron and hole spin, 
which corresponds to $s=N_e/2$ electrons and
$k=N_h/2$ holes having spin up and down, respectively. The
contribution of the other terms is small and vanishes in the
thermodynamic limit. Let us emphasize that for all results
presented below the maximum statistical error is about $5\%$,  
which is sufficient for the present analysis. Note that this accuracy can be
achieved at an acceptable cost of computer time. Of course, the  
error can be systematically reduced by increasing the length of
the Monte Carlo run.

\section{Simulation results}\label{simulations}
We now apply the theoretical scheme developed in the preceding
sections to a partially ionized dense electron-hole plasma.
We will be interested in strong Coulomb correlation effects such as
bound state (excitons, bi-excitons, clusters), their modification
by the surrounding plasma and their eventual breakup at high densities
due to pressure ionization (Mott effect). Beyond the Mott density, we
expect the possibility of hole crystallization if the hole mass is
sufficiently large \cite{Bonitz_PRL05}. To detect these effects, we have
extended our first-principle DPIMC simulations to a large range  
of mass ratios, temperatures and densities.
Below, the density of the two-component plasma is characterized by the
Brueckner parameter, $r_s=d/a_B$, defined as the ratio of the mean distance
between particles $d=[3/4\pi (n_e+n_h)]^{1/3}$ and the exciton Bohr
radius $a_B=\hbar^2\varepsilon/e^2 m_r$, where $n_e$ 
and $n_h$ are electron and hole
densities, respectively, and $\varepsilon$ 
is the background dielectric constant. In what follows 
we compute and discuss the spatial particle configurations, the pair
distribution functions, static structure factors,
fractions of electrons and holes in bound states, the internal energy and
the equation of state.

\subsection{Electron hole plasma with small/large mass ratio $M$}
\begin{figure}[t]
\includegraphics[width=6cm,clip=true]{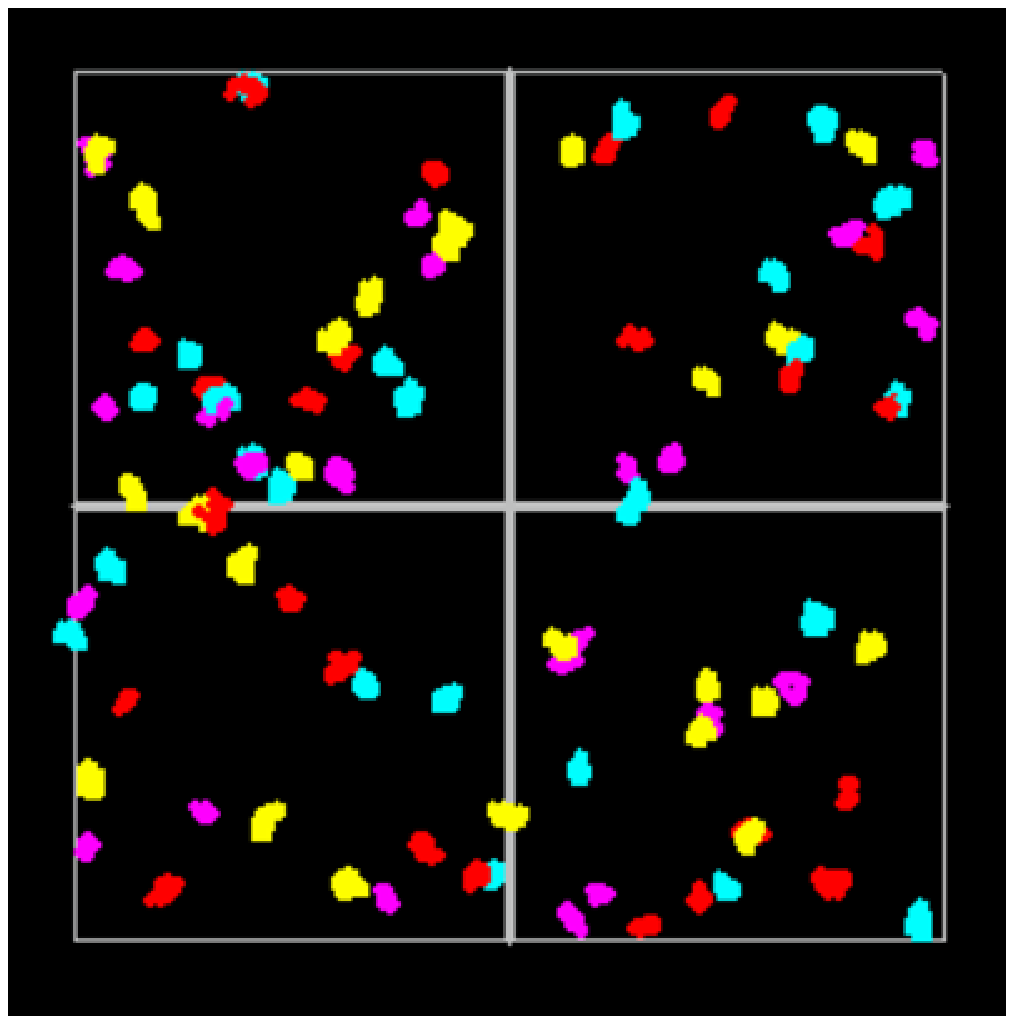}
\hspace{.5cm}
\includegraphics[width=6cm,clip=true]{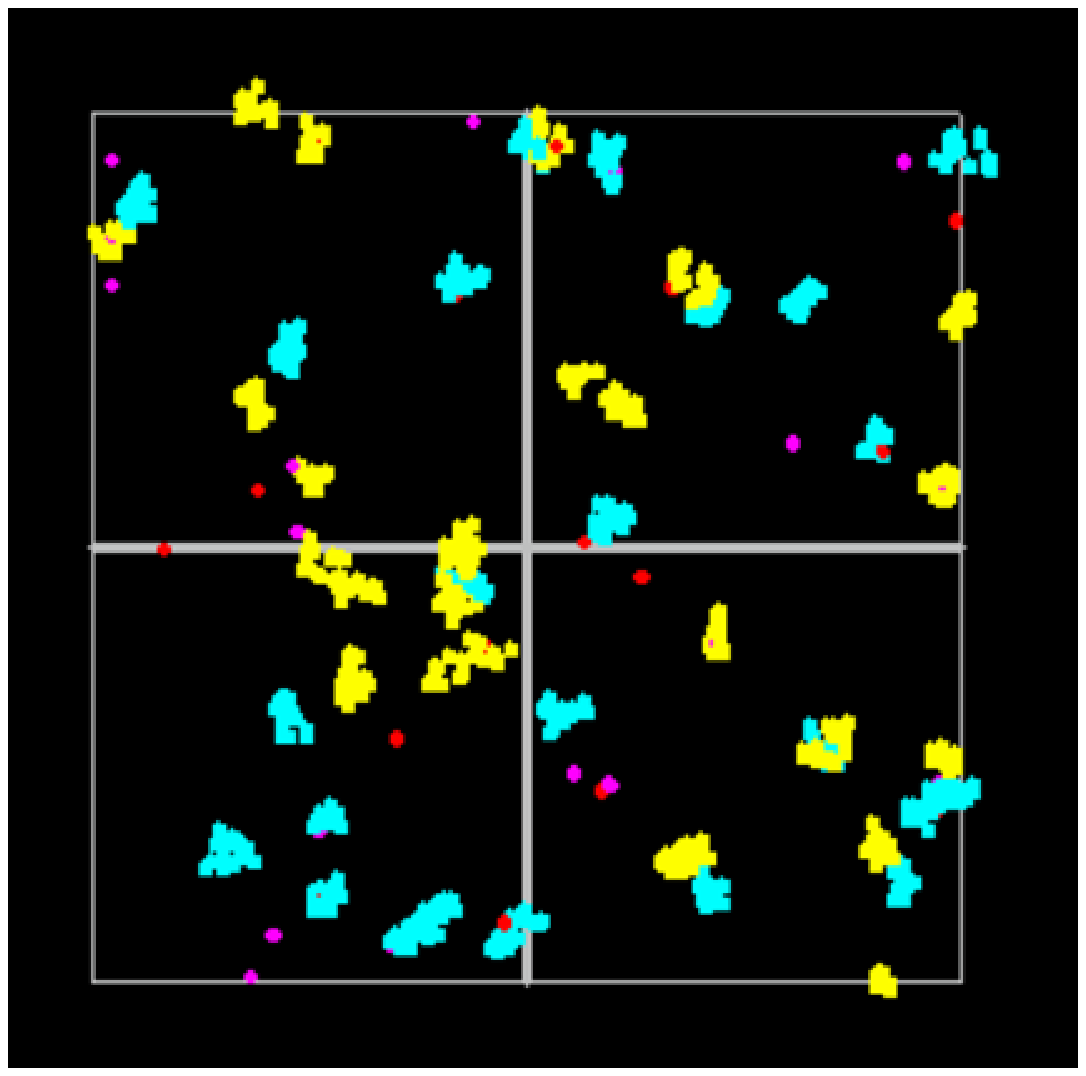}\\[0.4cm]
\includegraphics[width=6cm,clip=true]{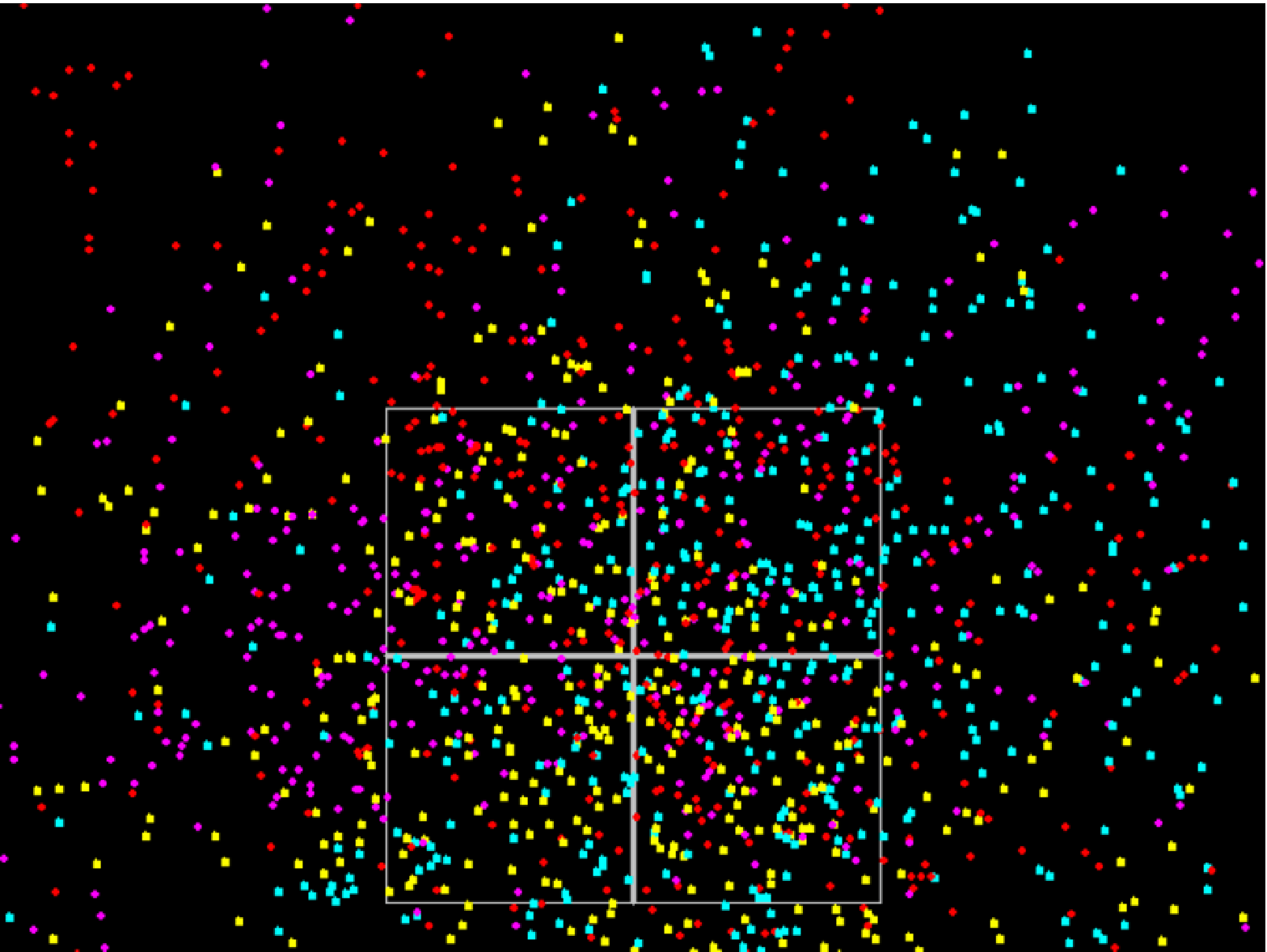}
\hspace{.5cm}
\includegraphics[width=6cm,clip=true]{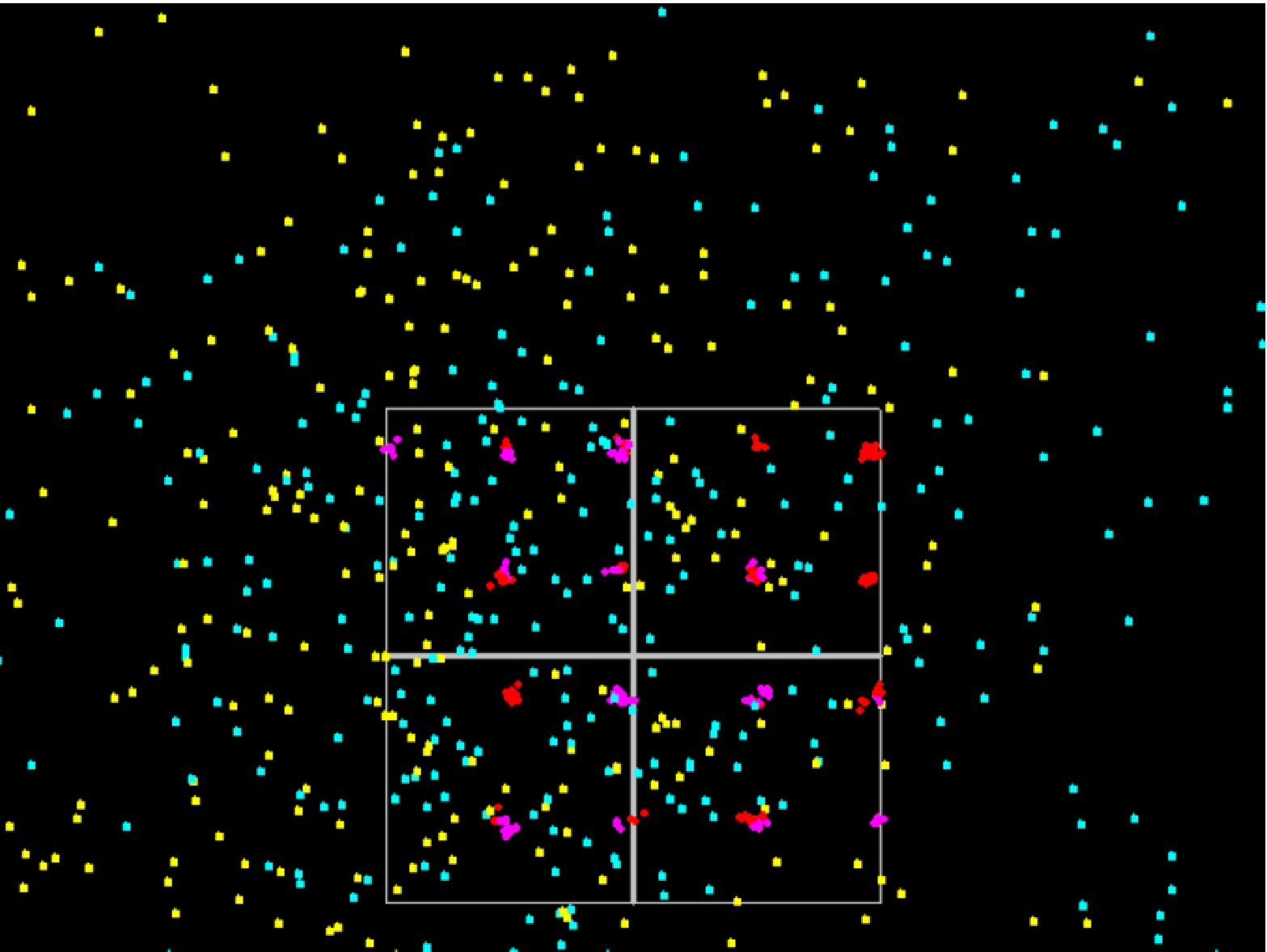}
\caption{(Color) Snapshots of the electron-hole plasma configuration in the simulation box for two mass 
ratios, left (right) column: $M=1, T=100K$ ($M=952$, T=50K) and two densities corresponding to, upper (lower) row: $r_s=10$ 
($r_s=0.33$).
Spin up and spin down electrons (holes) are
marked by yellow and blue (red and pink) clouds of dots, respectively, and 
the Monte Carlo cell is given by the gray grid lines (PBC were used).
\label{newfig3}
}
\end{figure}

\begin{figure}[t]
\includegraphics[width=10cm,clip=true]{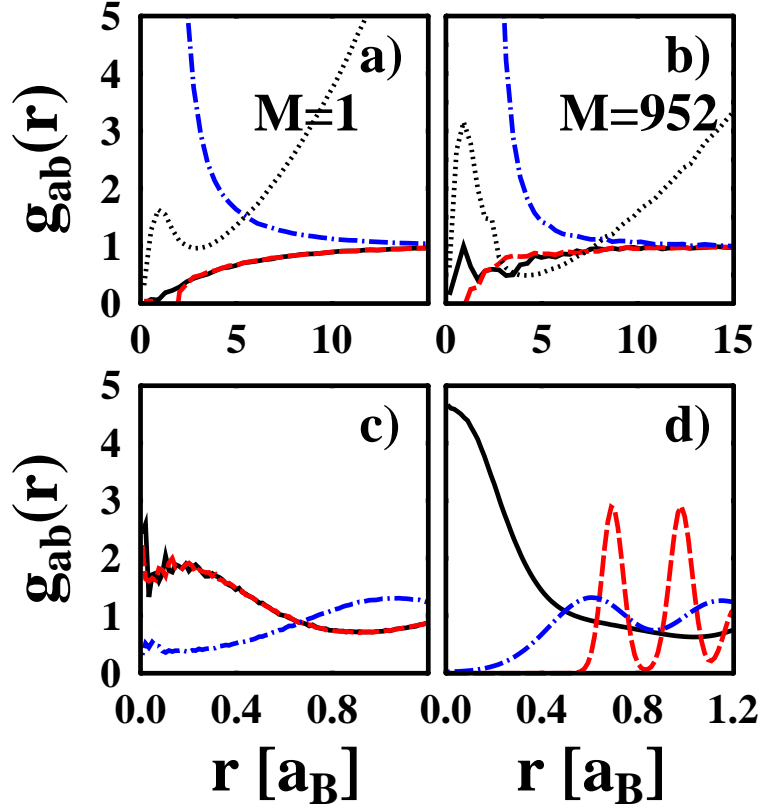}
\caption{(Color online) Pair distribution functions (PDF) for the four combinations of mass ratio and density shown in 
Fig.~\ref{newfig3}, i.e., Fig. a): $M=1$, $r_s=10$, b): $M=952$, $r_s=10$, c): $M=1$, $r_s=0.33$, d): $M=952$, $r_s=0.33$.
Full black lines: e-e PDF, dashed red lines: h-h PDF, dash-dotted blue lines: e-h-PDF, dotted black 
line: $r_{eh}^2 g_{eh}(r_{eh})$. Distances are in units of the exciton Bohr radius $a_B$, note the different scales in the 
upper and lower row.
\label{newfig4}
}
\end{figure}

Let us first discuss Coulomb correlations in two-component plasmas
for the limiting cases of small and very large mass ratios. 
In Figs.~\ref{newfig3},\ref{newfig4} we show data 
for $M=1$ (positronium, left columns in the two figures) and $M=952$ (right columns, illustrating the situation 
typical for hydrogen and plasmas of other chemical species). In addition, we consider the cases of 
small and large density, corresponding to $r_s=10$ (upper row) and $r_s=0.33$, (lower row in Figs.~\ref{newfig3} and~\ref{newfig4}) respectively, and temperatures well below the 
the exciton binding energy $E_B=e^2/2\varepsilon a_B$. 
To allow for a direct comparison 
with the Tm[Te,Se] system  below, we fix the effective electron mass to 
$m_e=2.1 m_0$ ($m_0$ is the free electron mass) and the background dielectric 
constant to $\varepsilon = 25$ which leads to a binding energy 
$E_B/k_B=517 K$ being two orders of 
magnitude smaller than for positronium and hydrogen.

In Fig.~\ref{newfig3} we present typical 
spin-resolved ``snapshots'' of the electron-hole state 
in the simulation box for small and large $M$ (left and right columns, respectively). 
One clearly sees the influence of the hole mass on the particle probability 
distribution (corresponding to the DeBroglie wavelength): In the left part, 
electron and hole ``clouds'' have the same 
size, whereas in the right part, holes have practically shrunk to dots (negligible size compared to the interparticle 
distance). 
Let us first concentrate on the low-density limit, $r_s=10$, upper row in Fig.~\ref{newfig3}. At this density Coulomb 
correlations are strong enough to give rise to bound states - excitons. They are clearly visible in the 
snapshots from pairwise clustering of electrons and hole ``clouds´´ for both, small and large mass ratios. Occasionally, 
also clusters of three or four particles are seen which correspond to trions (exciton ions) and bi-excitons, respectively.

A quantitative analysis of the behavior is obtained by computing the pair distribution functions (PDF) which 
are shown in the upper row of Fig.~\ref{newfig4}. They are computed in the simulations 
from the density operator according to 
\begin{equation} \label{gab-rho}
g_{ab}(r) = \frac{N_e!N_h!}{Z} \sum_{\sigma}
\int\limits_{V} dq \,\delta(r_{1,a}-q_{1,a})\,
\delta(r_{2,b}-q_{2,b}) \,\rho(q,\sigma;\beta)\,.
\end{equation}
For both mass ratios we observe a clear peak of $r^2g_{eh}$ at about 
$r_{eh}=1a_B$, corresponding to excitons. For the large mass ratio, Fig. \ref{newfig4}b), the peak is substantially 
higher which is a consequence of the increase 
of the binding energy by a factor two 
compared to the case $M=1$ (the reduced 
mass increases from $m_e/2$ to $m_e$) which stabilizes 
the bound states. This trend is also seen in the 
snapshots (upper panel): The fraction of 
electron ``clouds'' closely attached to holes is significantly higher in the right figure. Further, the h-h PDF for 
the large mass ratio, Fig. \ref{newfig4}b), show signatures of bi-exciton formation with a peak distance of 
about $1.4a_B$. Also the e-e PDF shows peaks, one at lower distances, correspponding to electrons with different 
spin projections located between the holes and a weaker pronounced one at larger distances. These peaks are not seen 
for $M=1$.

Let us now turn to the limit of high densities, lower row in Figs.~\ref{newfig3} and~\ref{newfig4}. Here the influence of the 
mass ratio is even more dramatic, leading to a qualitative change of the plasma behavior. 
While the electrons are practically delocalized over the whole 
simulation volume for both $M$, the behavior of the holes changes 
from delocalized (small mass ratio; lower left part of Fig.~\ref{newfig3}) to fully localized 
(large mass ratio; lower right part). 
Obviously, in both cases no bound states exist. Instead, we observe 
a Fermi gas-like state of electrons and holes, at small $M$, 
and a hole crystal which is 
embedded into an electron Fermi gas, at large $M$. 
This interpretation is confirmed by the 
behavior of the PDF (lower row of Fig.~\ref{newfig4}). 
They are almost structureless at small $M$. In contrast, at large $M$, 
pronounced peaks at finite $r$ are visible in the hole-hole PDF being  
typical for a crystalline structure~\cite{Bonitz_PRL05}.

Thus, Figs.~\ref{newfig3} and~\ref{newfig4} allow us to conclude that
in order to observe a crystal of charged particles (hole crystal) 
in a two-component quantum plasma, three requirements have to be fulfilled: 
(i) sufficiently low temperature, 
(i) sufficiently high density (which causes pressure ionization of 
the bound states) and (iii) a large mass ratio. 
Below, these requirements will be studied more in detail.

The most interesting question is the role of the mass ratio. 
Crystals of ions (e.g. nuclei of carbon and oxygen) in the presence 
of a quantum electron gas are commonly accepted to exist in astrophysical objects 
such as White Dwarf stars~\cite{segretain}, where the mass ratio is of the order $M\sim 10^4$. 
But ion crystallization is expected to be possible 
also for much smaller mass ratios: proton crystallization ($M=1836$) 
in dense hydrogen has been found in our previous 
simulations~\cite{filinov-etal.00jetpl}, see also Ref.~\cite{militzer06}, and 
we have also found crystals of alpha partilces in pure 
helium~\cite{Bonitz_PRL05}. Figs.~\ref{newfig3} and~\ref{newfig4} indicate that crystallization of holes  
should be possible even for $M$ below 1000. 
Two-component plasmas with $M<1000$ exist in condensed matter systems, such as semiconductors. 
In most traditional semiconductors, however, 
typical values of $M$ are $3\ldots 20$. 
There have been predictions of the possibility of $M\sim 100$ in CuCl or 
Bi-Sb alloys under pressure, e.g.~\cite{rice68,abrikosov78}, 
but without experimental evidence so far. 

\subsection{Thermodynamic properties of electron hole plasmas with $M=40$}
The largest mass ratios in condensed matter systems 
were experimentally reported  by Wachter and co-workers \cite{wachter}
for the intermediate valent $\rm Tm[Se_xTe_{1-x}]$ alloys under pressure. 
In these materials f-d-hybridization 
provides a narrow dispersive f-valence band and, as a
consequence, a large effective hole mass of the order of
50-100 (bare) electron masses.  
This system is of particular interest because 
of the long life time of the electron-hole plasma. Moreover 
the mass ratio is close to the predicted 
critical value of $M\approx 80$ 
for hole crystallization~\cite{Bonitz_PRL05}. 
\begin{figure}[b]
\includegraphics[width=8cm,clip=true]{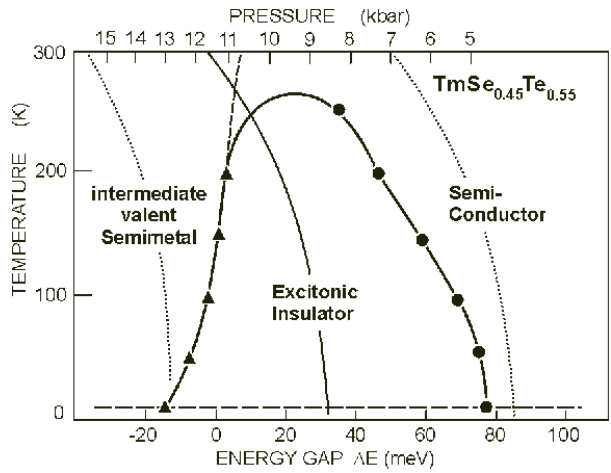}
\caption{Excitonic phase diagram of ${\rm TmSe_{0.45}Te_{0.55}}$ taken from Ref.~\cite{wachter}.
Measured points are designated by symbols. ``Isobars'' in
the semi-conducting and semi-metallic phase are marked as dotted lines,
whereas the ``isobar'' entering the excitonic phase is denoted
by a full line. The lower abscissa gives the corresponding
energy gap $\Delta E$ (here negative values refer to the
metallic state). The left boundary (at high pressure) of 
the exciton-rich phase corresponds to the pressure ionization (Mott effect).
For more details see Ref.~\cite{wachter,BF06}. 
\label{newfig5}
}
\end{figure}

$\rm TmSe_{0.45}Te_{0.55}$ at ambient conditions 
is an indirect semiconductor with a gap of
$E_{\Delta}=130$~meV. An excitonic level has been observed 
with $E_B\simeq 50-70$~meV below the bottom of the d-band.
Applying pressure the gap can be tuned (and even closed), and the
material is speculated to realize in the pressure region between 5
and 11~kbar an excitonic insulator, at least at very 
low temperatures~\cite{BF06}, the search for which has been
run for a long time~\cite{rice68} (see the experimental
phase diagram in Fig.~\ref{newfig5}). 
A necessary pre-condition is the existence of
a large number density of (up to $10^{20}-10^{21}$ per cc) excitons of
intermediate size (in order to avoid too strong overlap of the
excitonic bound states)~\cite{wachter}. 
For pressures exceeding 11~kbar exciton ionization 
has been observed as a result of the Mott effect. This is the region 
in phase space where hole crystallization should occur.

In the following, we analyze this interesting system more in detail
performing direct PIMC simulations within the parabolic band approximation. 
While this is certainly a strongly simplified model, 
we expect that the main trends, related to density 
and temperature variation, will be reproduced.
The simulations were performed for an e-h plasma 
with $m_e=2.1 m_0$ ,$m_h=80m_0$, and $\varepsilon = 25$. 
Several characteristic temperatures within the interval of 
the measurements by Wachter et al. \cite{wachter} 
($T \lesssim 300 K$) are chosen, which are well below the
exciton binding energy.

\subsubsection{Pressure and internal energy}
\begin{figure}[t]
\includegraphics[width=8cm,clip=true]{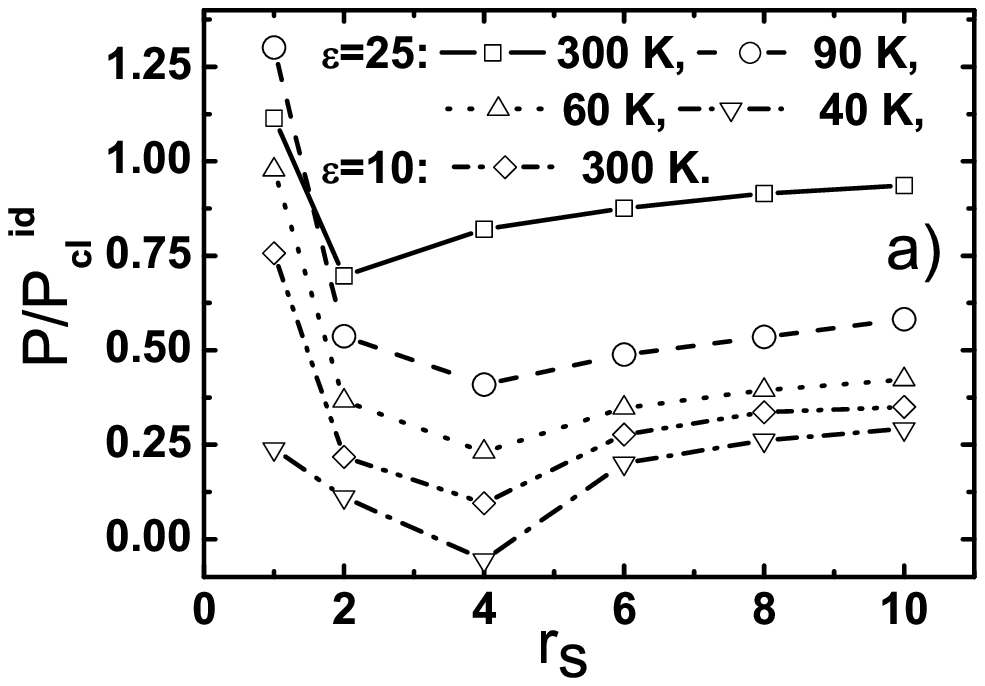}
\includegraphics[width=8cm,clip=true]{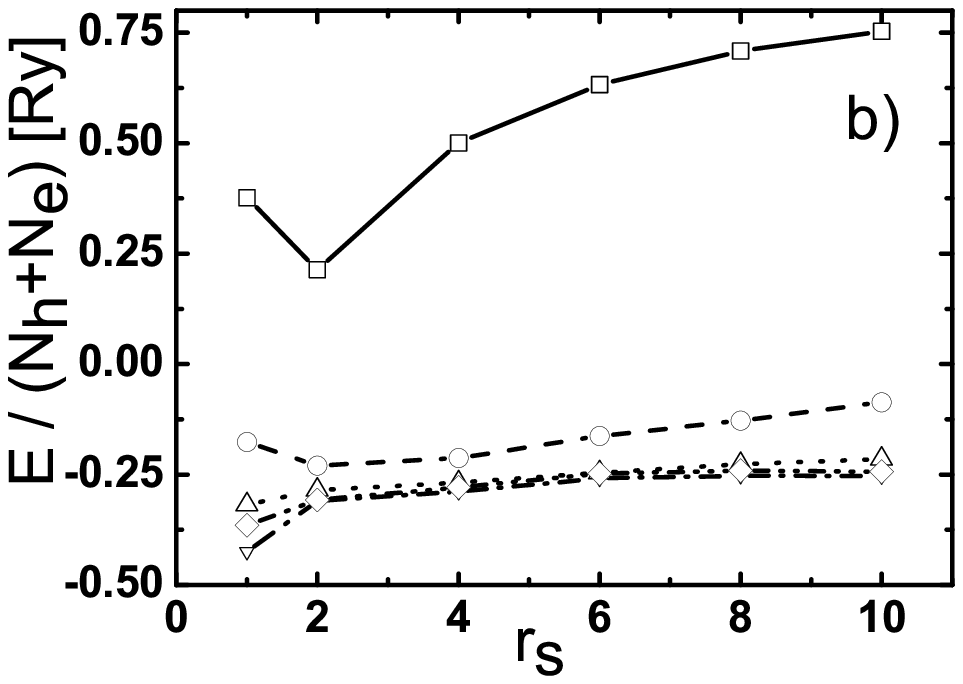}
\caption{Isotherms of pressure and internal energy versus Brueckner parameter ($r_s\sim n^{-1/3}$)  
for an e-h-plasma with $m_e=2.1m_0$, $m_h=80m_0$ and $\varepsilon=25$. For comparison, also results 
for $\varepsilon=10$ at $T=300$K are shown.
Pressure is in units of the classical ideal pressure, $P^{id}_{cl}=(n_e+n_h)k_BT$.}
\label{pie}
\end{figure}

We first consider the behavior of pressure $p$ and internal energy $E$ 
which are determined in our simulations exploiting 
formulas~(\ref{eos}) and (\ref{energy}), derived in the App.~\ref{theory_b}.
Figure~\ref{pie} shows $p$ and $E$ versus the Brueckner 
parameter (i.e. versus density to the power $-1/3$) at several 
fixed temperatures. At low densities and 
high temperature ($300$K), $p$ and $E$ 
reflect the classical ideal gas behavior. 
Reduction of density or/and temperature lead 
to a significant deviation from this behavior: Now
$p$ and $E$ decrease due to the (overall attractive) 
Coulomb interaction in the plasma and, in particular, 
due to bound state formation of excitons and 
biexcitons, reaching a minimum around $r_s \sim 2\dots 4$. 
For higher densities, $p$ and $E$ start to increase again monotonically. 
This is triggered by quantum effects - the 
plasma behaves as a Fermi gas. 
At the same time, bound states 
are expected to break up as a consequence of the 
Mott effect. 
The behavior of bound states will be verified below from the snapshots 
and pair distribution functions. 
The negative values of $p$ observed at the lowest 
temperature point towards an instability 
of the homogeneous plasma state against formation 
of droplets or clusters. Electron-hole droplet 
formation in semiconductors is well established and 
was observed experimentally three decades ago~\cite{keldysh}. This effect 
is similar to the so-called plasma phase transition discussed by many authors 
for dense hydrogen and other 
plasmas (see~\cite{norman_starostin,Pade85,sbt95,BeEb99,filinov-etal.01jetpl} 
and references therein).

\subsubsection{Particle configurations}
\begin{figure}[t]
\includegraphics[width=4.6cm,clip=true]{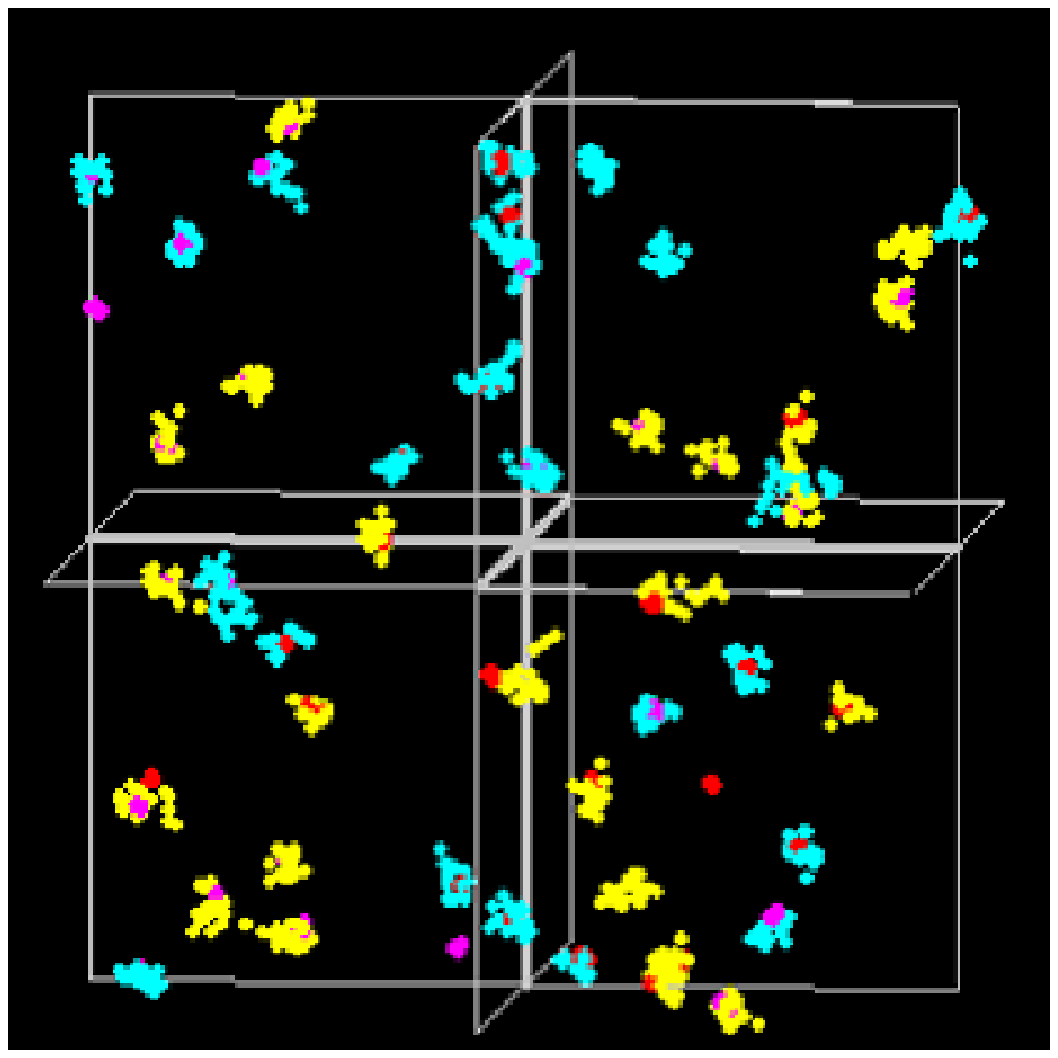}
\hspace{.5cm}
\includegraphics[width=4.6cm,clip=true]{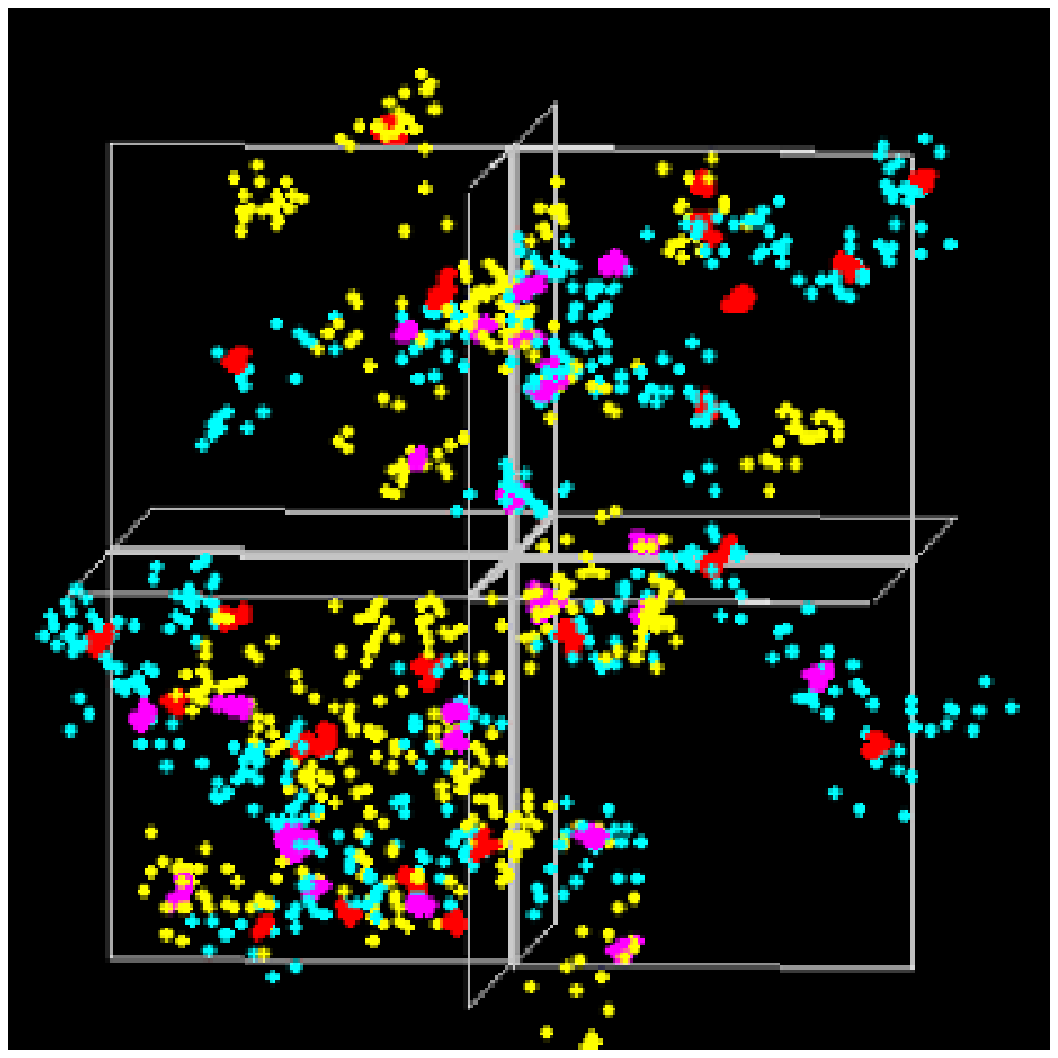}
\hspace{.5cm}
\includegraphics[width=4.6cm,clip=true]{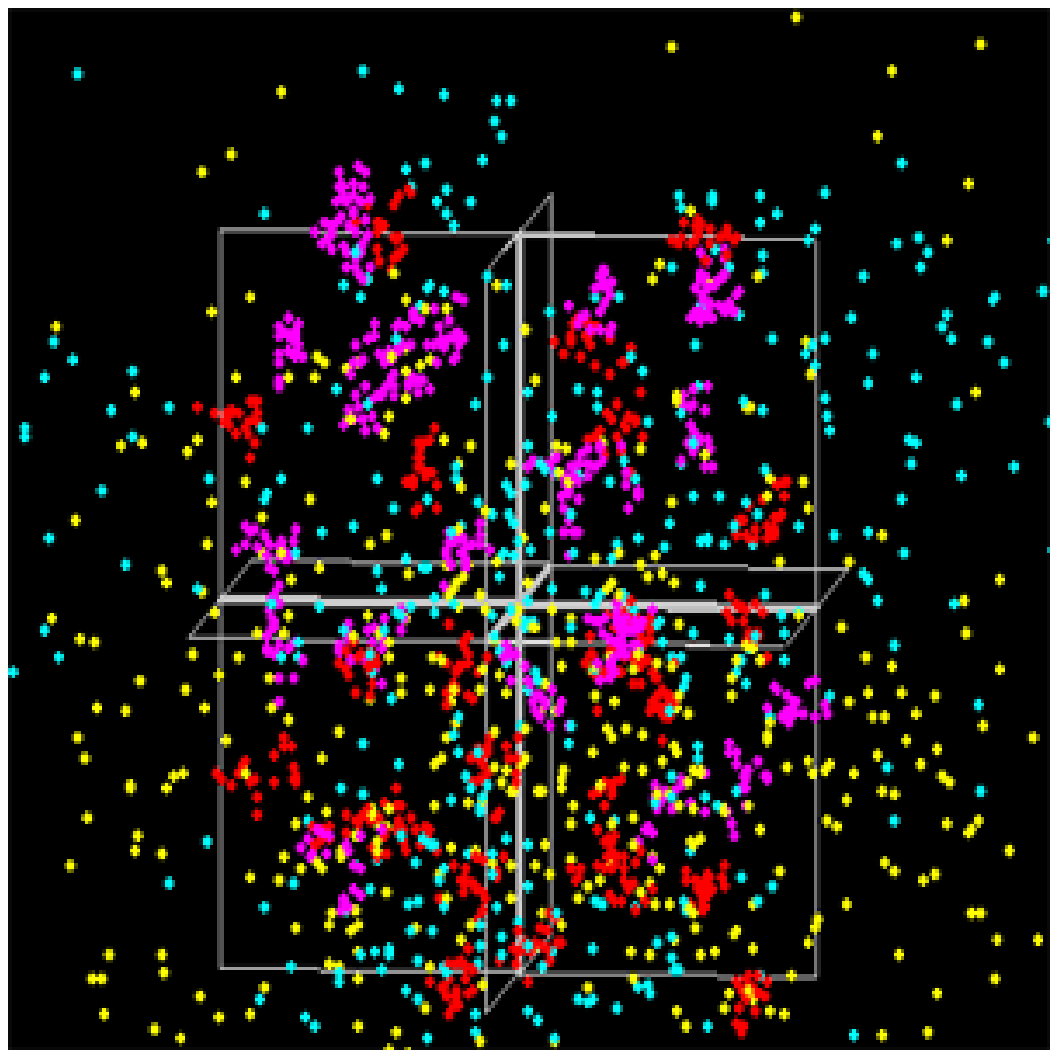}\\[0.4cm]
\includegraphics[width=4.6cm,clip=true]{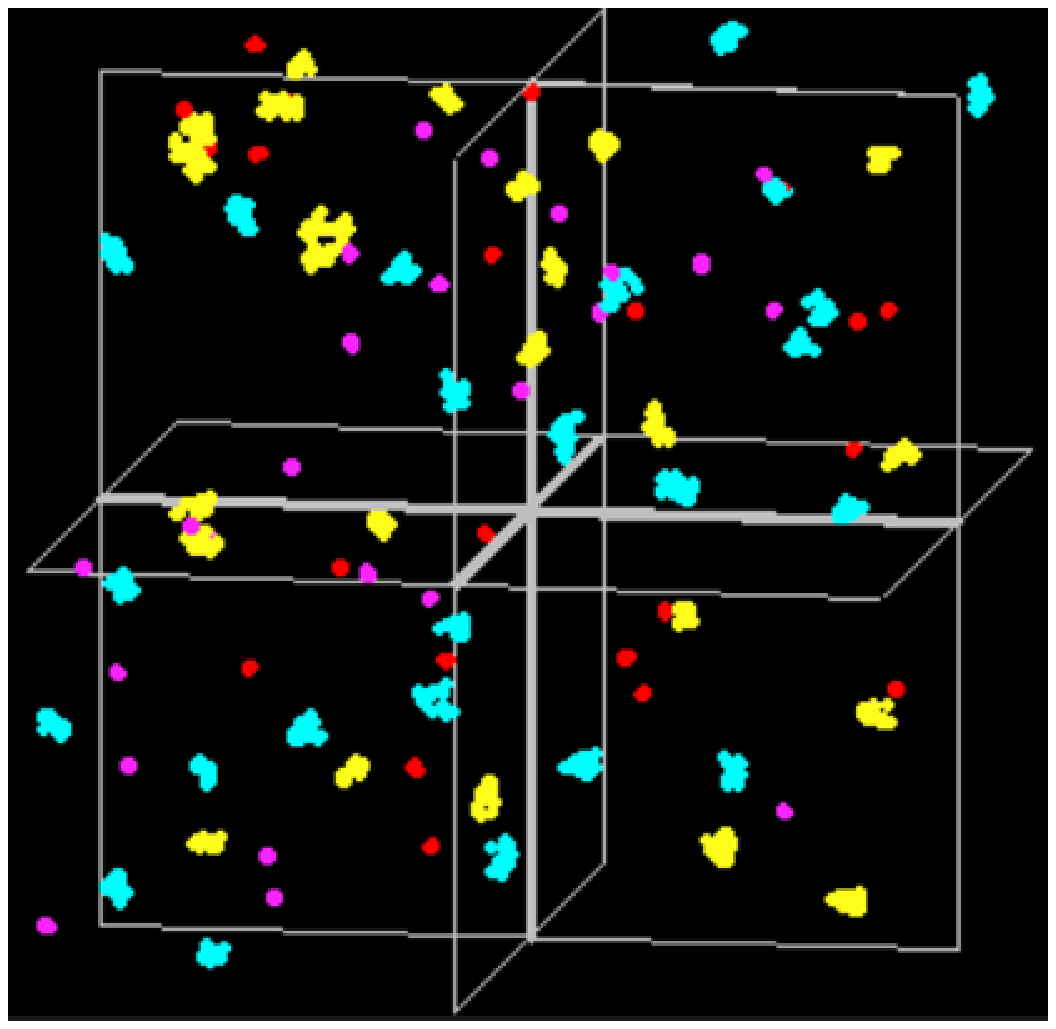}
\hspace{.5cm}
\includegraphics[width=4.6cm,clip=true]{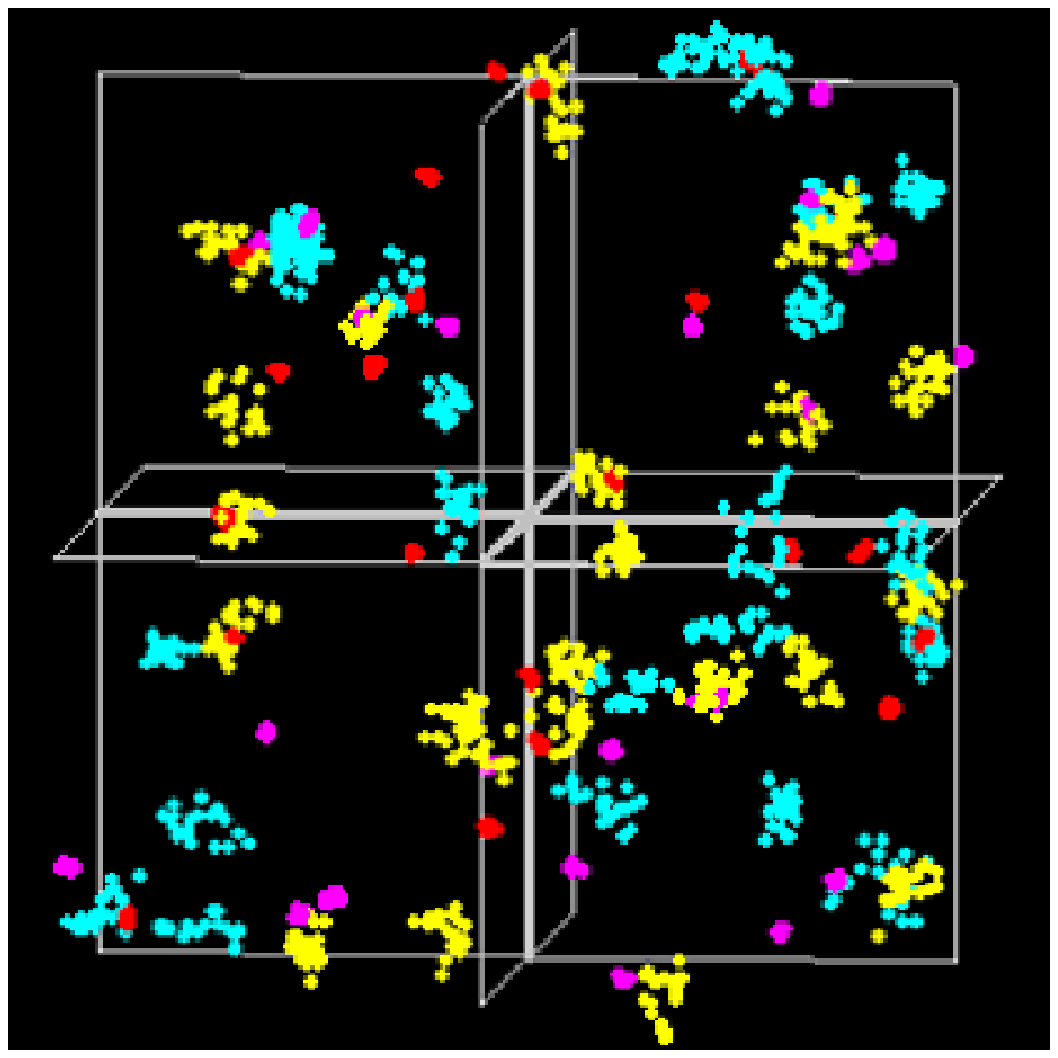}
\hspace{.5cm}
\includegraphics[width=4.6cm,clip=true]{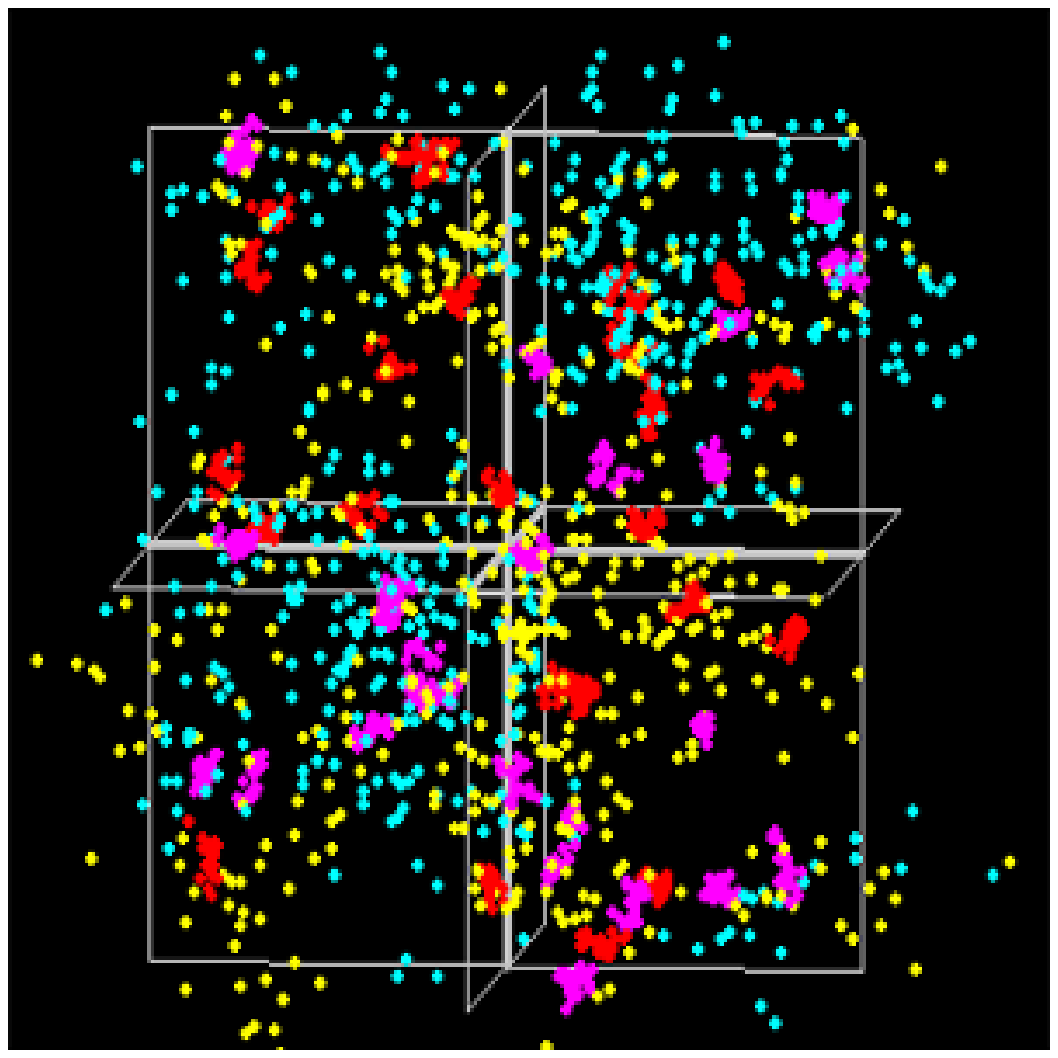}
\caption{(Color) Snapshots of particle configurations
at three particle densities corresponding to $r_s=10$ (left), $r_s=4$ (center),
and $r_s=1$ (right). Spin up and spin down electrons (holes) are
marked by yellow and blue (red and pink) clouds of dots, and 
the Monte Carlo cell is given by the gray grid lines (PBC were used).
Upper and lower panels show results at low ($T = 50 K$) and high ($T = 200 K$)
temperatures,  respectively. 
\label{Snpsht}
}
\end{figure}

Figure~\ref{Snpsht} shows snapshots of the electron-hole configuration in the
simulation box at different temperatures and densities. According to the temperature
decomposition of the density matrix, each electron and hole is represented
by several beads. We used $n=20 $ beads, 
so for the upper panels, at $T=50$~K, the high-temperature density 
matrices are taken at a temperature of~$1000 K$,
being two times larger than $E_B$.
The spatial distribution of the beads of each quantum particle is
proportional to its spatial probability distribution. 
Figure~\ref{Snpsht} indicates that for $M\approx 40$ 
the typical size of the cloud of beads for electrons 
is several times larger than the one for the holes. 
Note, that in the present strongly correlated system, the 
extension of electron and hole probability densities maybe quite different
from the deBroglie wavelength which corresponds to the ideal case.

At low density ($r_s=10$, left column)
and temperature $T=50 K$, practically all holes are closely
covered by electron beads, which means that electrons and holes form bound states, and
the e-h plasma consists mainly of excitons. This interpretation
will again be supported by the
behavior of the pair distribution functions discussed below.
Raising the temperature at fixed density leads to a (temperature-induced)
ionization of the bound states. As a result we find a
substantial number of free electrons and holes in the simulations 
(the degree of ionization will be shown in Fig.~\ref{cmpexp} below).

For intermediate densities ($r_s=4$, middle column) we observe
the formation of bi-excitons and many-particle clusters 
at low temperatures ($T=50 K$).
Now the electron-hole system is strongly inhomogeneous. 
In this case the mean distance between particles $d$ 
is of the order of the electron wavelength. At $T=200 K $
bi-excitons and many-particle clusters are absent due to 
thermal break up of exciton-exciton bound states.

At high density ($r_s=1$) the electron wavelength
exceeds the mean inter-particle distance $d$ and even the size of the Monte Carlo cell used in the simulations, which is seen
by the large extension of the electron beads clouds. At the same time 
excitons become unstable because two electrons bound to neighboring holes
start to overlap allowing for electron tunneling from one exciton to another
(pressure ionization, Mott effect), and the system 
transforms into a plasma of electrons and holes. 
Since the hole wavelength is significantly smaller than the
electron wavelength (and may still be smaller than $d$), 
in a certain region of $r_s$-values the structure 
of the hole beads resembles a liquid state (lower right panel).

\subsubsection{Pair distribution functions and structure factors}
\begin{figure}[b]
\centerline{
\includegraphics[width=7.7cm,clip=true]{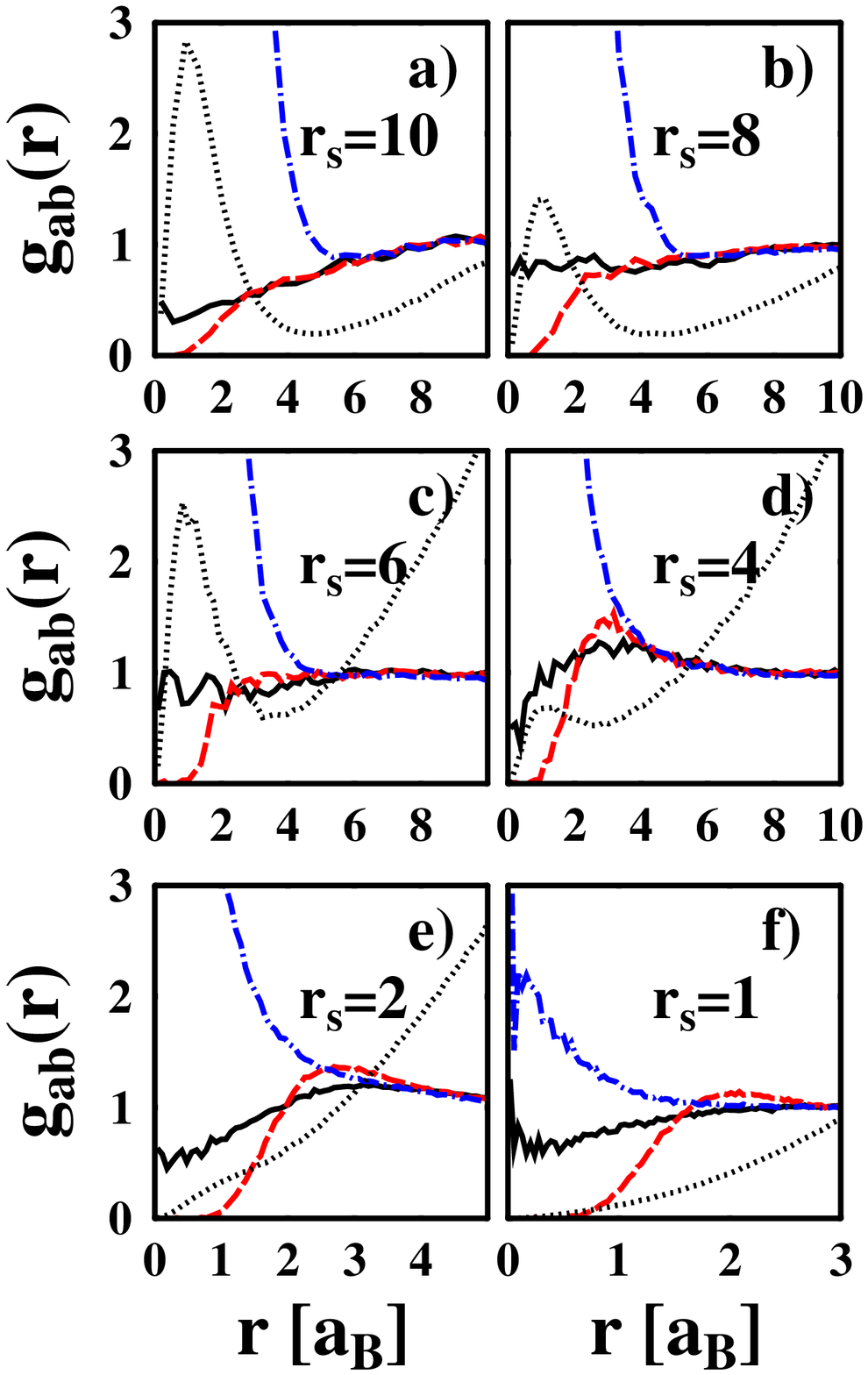}\hspace{0.5cm}
\includegraphics[width=6.8cm,clip=true]{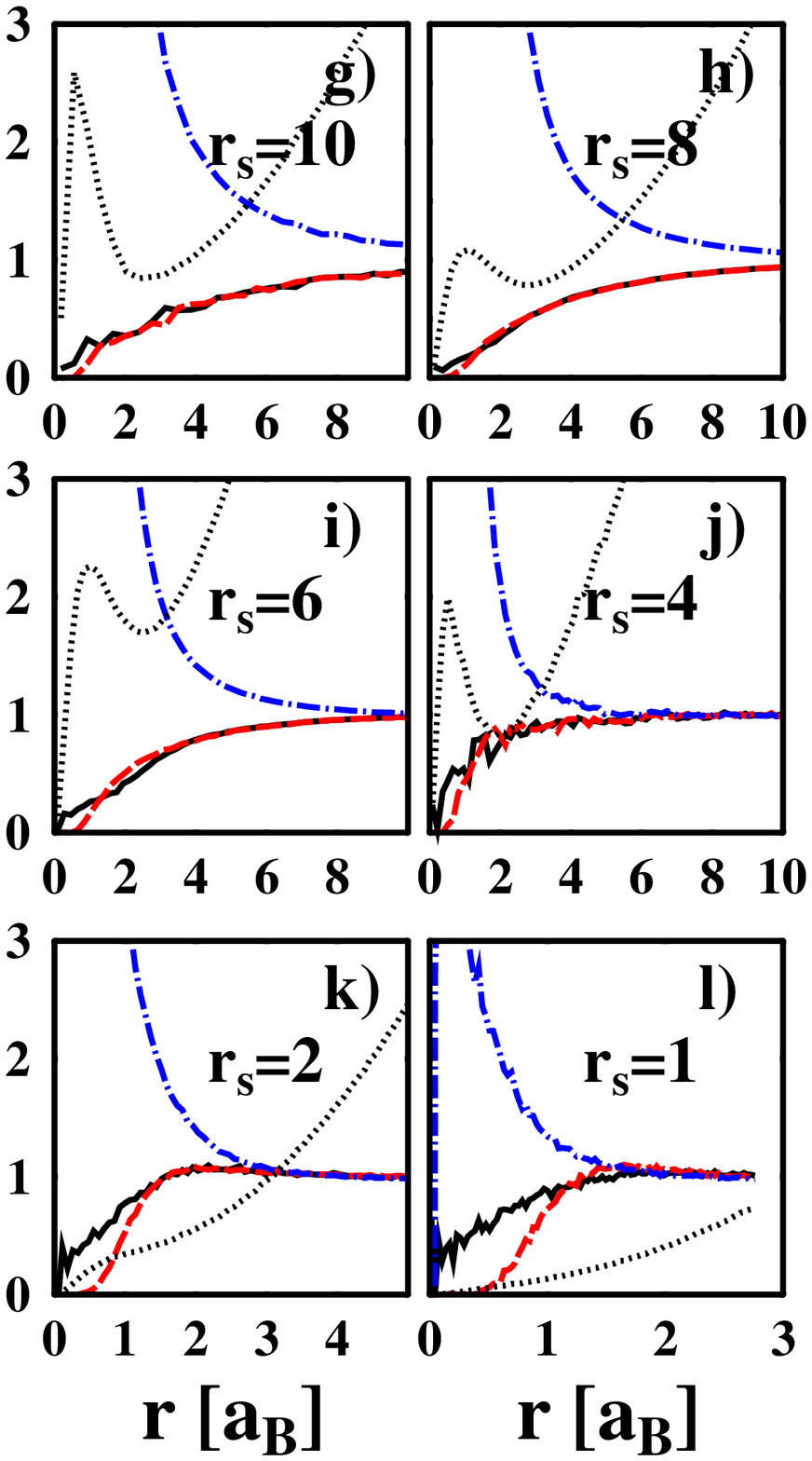}}
\caption{(Color online) Pair distribution functions $g_{ee}$ (black solid line),
$g_{hh}$ (red dashed line), and $g_{eh}$ (blue dot-dashed line) at
$T=50K$ (left two columns) and $T=200 K$ (right two columns). The
dotted lines show $r^2g_{eh}$ where, for better visibility of the
exciton peak, the data are divided by a factor $120$ in panels a and b, 
by a factor of $30$ in figures c, d, g, and h, and by a factor of 
$10$ in panels e, f, and i-l.}
 \label{geehh}
\end{figure}

Now let us discuss the behavior of the spin-averaged
electron-electron-, hole-hole- and electron-hole PDFs 
defined in Eq.~(\ref{gab-rho}).
Figure~\ref{geehh} shows the three functions $g_{ab}(r)$ versus the
inter-particle distance $r=r_{1,a}-r_{2,b}$ for the
same densities and temperatures as in Fig.~\ref{Snpsht}. 
Here no smoothening has been carried out,
i.e. the fluctuations of $g_{ee}, $ $g_{hh}$ and $g_{eh}$ at small $r$
reflect the magnitude of the statistical errors of our simulation.

\begin{figure}[h]
\centerline{
\includegraphics[width=7.5cm,clip=true]{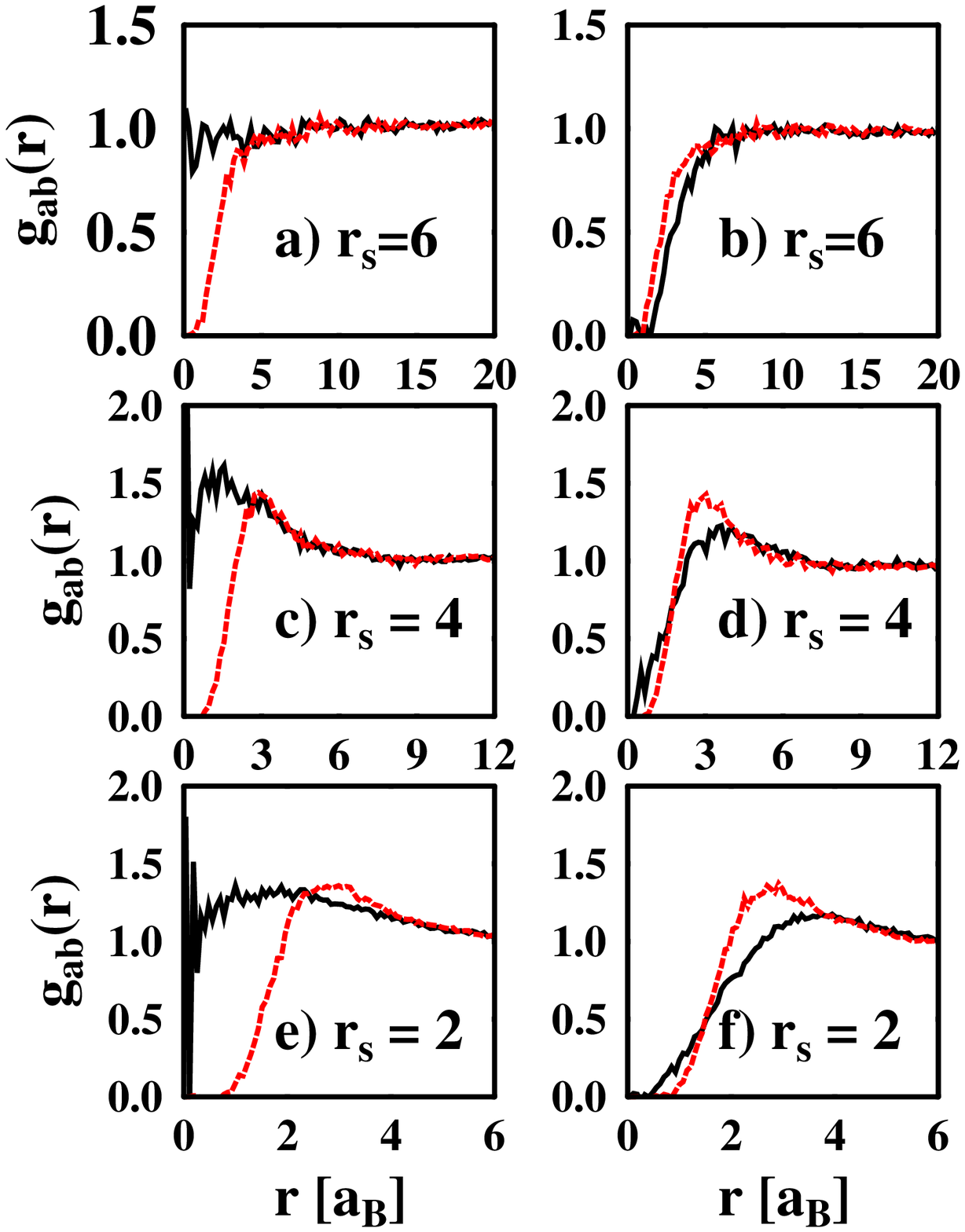}\hspace{.5cm}
\includegraphics[width=7.25cm,clip=true]{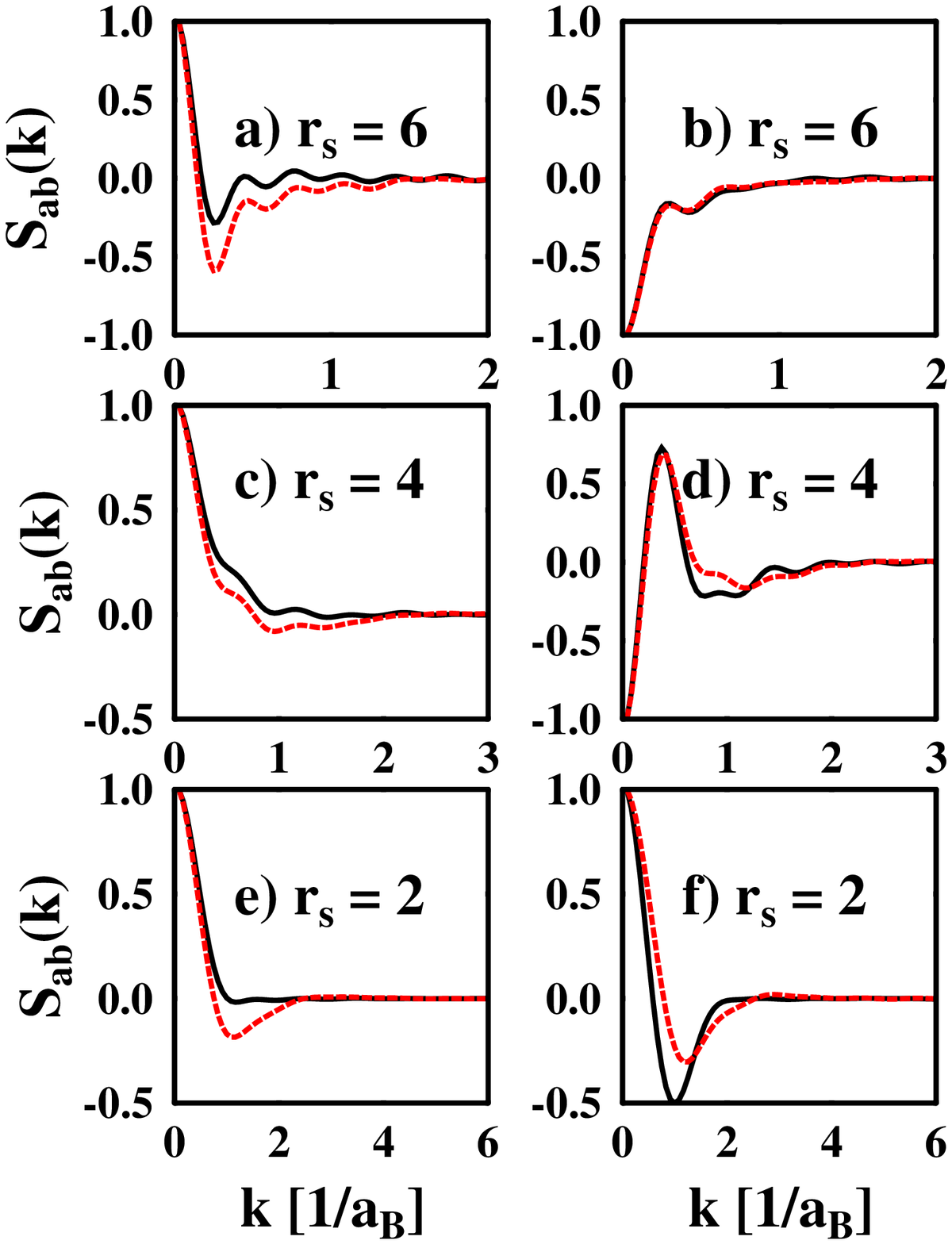}}
\caption{(Color online) Spin resolved pair distribution functions $g_{ab}(r)$ and
structure factor $S_{ab}(k)$ for electrons (solid black lines) and holes 
(dashed red lines) at $T=50$~K. Left (right) columns correspond to antiparallel (parallel)
spin projections of the particle pair.}
 \label{gseehh}
\end{figure}

As an effect of the Coulomb and Fermi (statistics) repulsions, 
$g_{ee}$ and $g_{hh}$ are suppressed
at small distances. Due to the large mass difference
the decay of the e-e correlations, however,
is essentially different from that of the h-h correlations. 
The asymptotic of $g_{ee}$ at small
distances is determined mainly by electrons with
opposite spin projections (since the strong Fermi repulsion is absent).
For these electron pairs the main contribution to the repulsion comes from the
effective quantum Kelbg potential (recall that it is finite at zero 
distance), allowing for tunneling of electrons up to zero separation. 
This is supported by the behavior of the spin dependent pair distribution
shown in Fig.~\ref{gseehh}. Of course, the tunnelling effect are 
much more pronounced for the lighter electrons.

The strong peak of $g_{eh}$ at low densities ($r_s=10\dots 8$) is caused by
excitons. This is confirmed by considering the function $r^2 g_{eh}(r)$ which
exhibits a pronounced maximum at about $1 a_B$, i.e. at the exciton 
Bohr radius. At these densities, the functions
$g_{ee}$ and $g_{hh}$ exhibit no peak structure,
i.e. there is no indication of formation of bound exciton-exciton complexes
(such as bi-excitons or electron-hole droplets). With increasing temperature
more and more excitons dissociate, and the maximum of $r^2 g_{eh}$ is reduced.

Varying, at low temperatures, the density over three orders of magnitude,
the maximum of $r^2 g_{eh}$ is suppressed and finally vanishes.
At about $r_s=6$, there is evidence that recombination into bi-excitons
takes place. At the same time, the position of the maximum in
$r^2 g_{eh}$ shifts from 1~$a_B$ to 2-3~$a_B$ approximately, indicating
the increasing radii of the bound states. This scenario is confirmed by the
behavior of $g_{ee}$ and $g_{hh}$:
For intermediate densities ($2< r_s < 6$), they
exhibit distinct peaks at larger $r$,
pointing towards the formation of bi-excitons and
many-particle clusters. For $r_s \lesssim 1$, the fraction of excitons
is further reduced  due to many-body effects (pressure ionization),
and bi-excitons and many-particle clusters vanish.

Finally let us relate the width of the peak of $g_{eh}$ to
the extension of the ground state wavefunction ($\sim 1 a_B$).
At low densities ($r_s > 6$) the peak of $g_{eh}$ is rather broad,
indicating the population of excited states, while at high densities
($r_s \sim 1$) the width of the lowest peak
in $g_{eh}$ becomes significantly smaller than the corresponding
width of the ground-state peak. At the same time the
hole-hole pair distribution function reveals an ordering of the
holes into a fluid-like state.

\begin{figure}[h]
\centerline{
\includegraphics[width=7.4cm,clip=true]{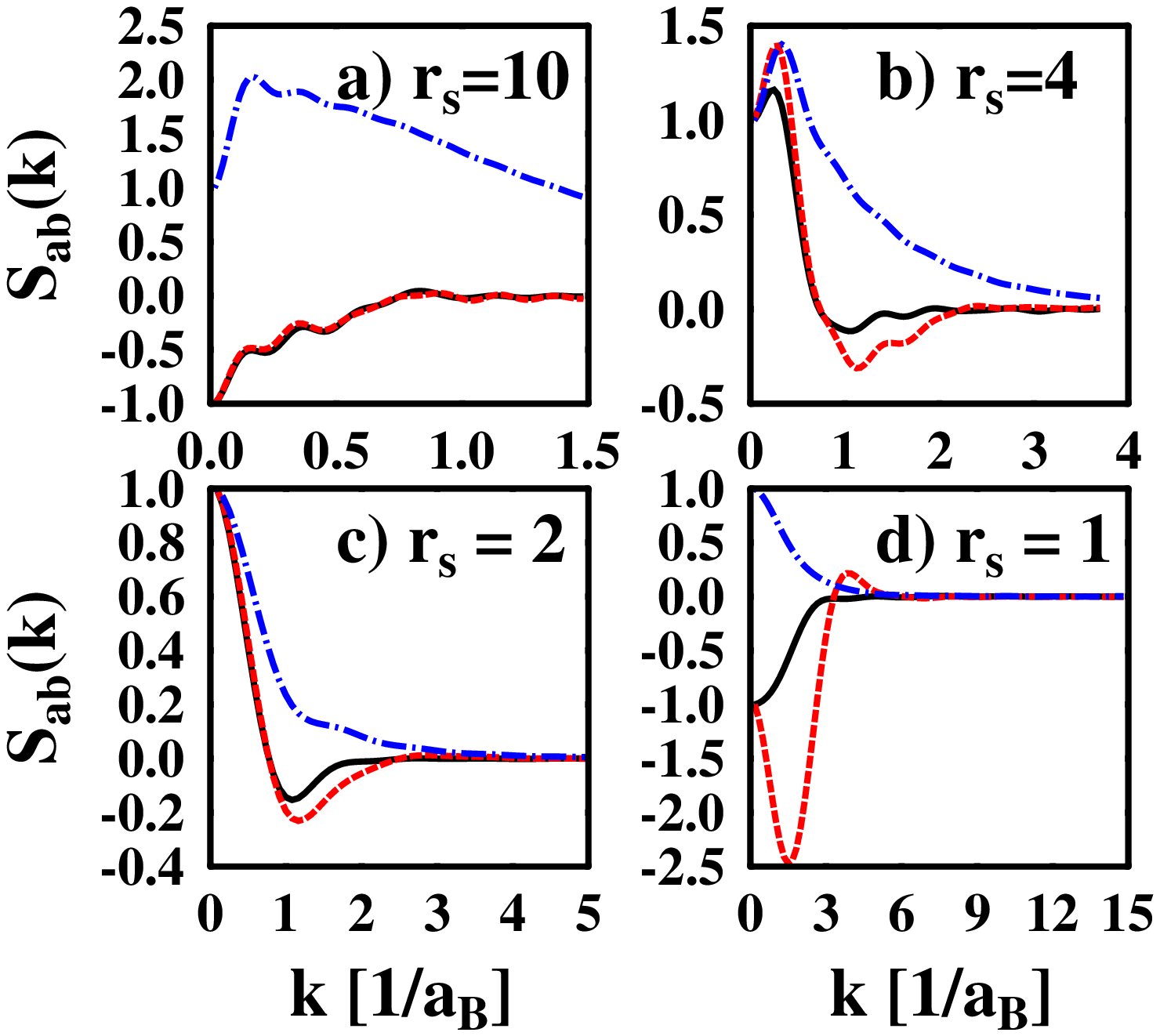}\hspace*{.5cm}
\includegraphics[width=7.4cm,clip=true]{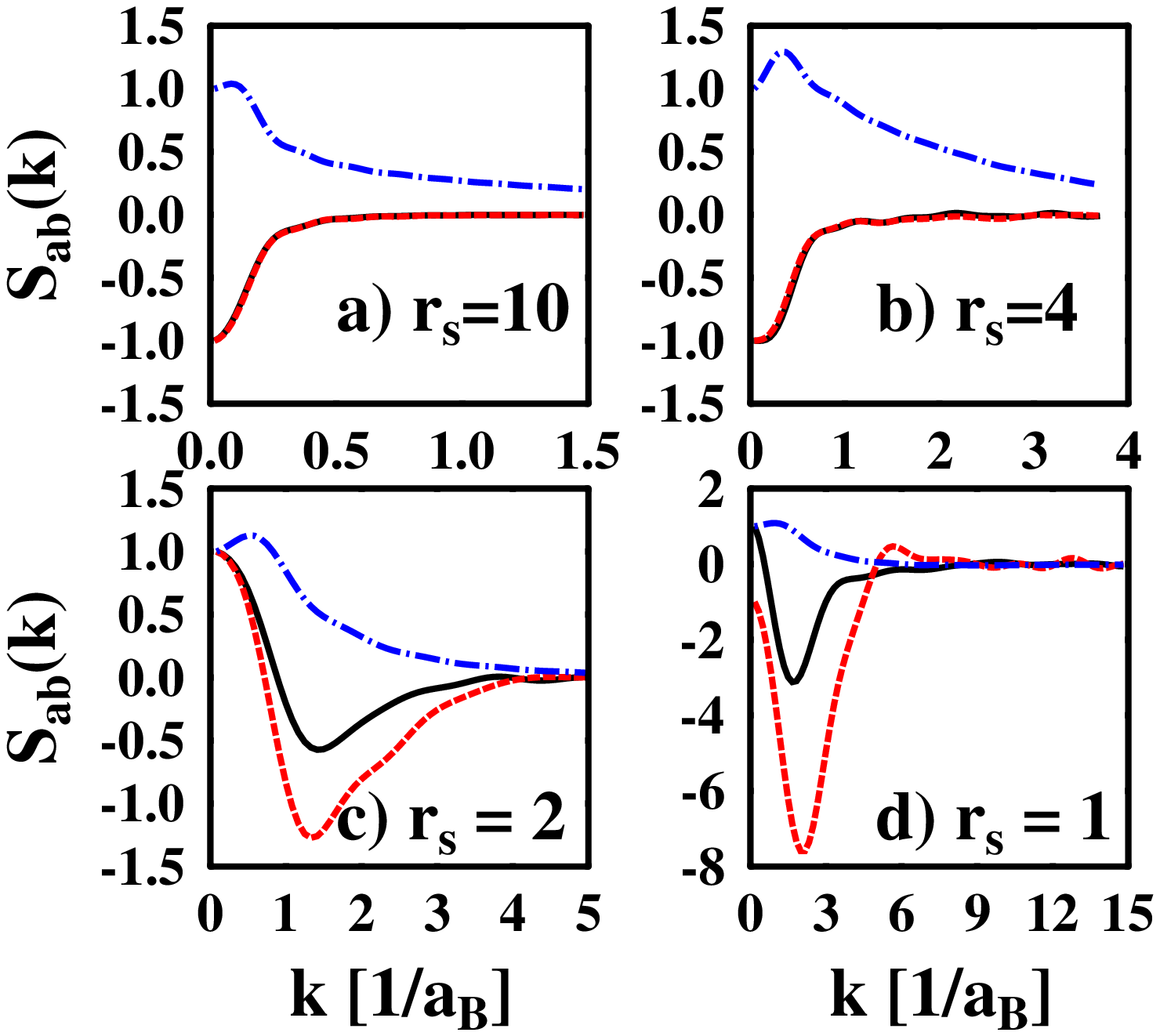}}
\caption{(Color online) Static structure factor $S_{ee}$ (black solid line),
$S_{hh}$ (red dashed line), and $S_{eh}$ (blue dot-dashed line) at $T=50 K$ (left columns)
and $T=200 K$ (right columns).} \label{seehh1}
\end{figure}

Figure~\ref{seehh1} shows the variation of the static charge structure factor,
defined in reciprocal $k$-space as
\begin{eqnarray}
S_{ab}(k) &=& \frac{\int_{0}^{\infty}dr
r^2(g_{ab}(r)-1)\sin(kr)/(kr)}{|\int_{0}^{\infty}dr
r^2(g_{ab}(r)-1)|}\,. 
\label{sab}
\end{eqnarray}
According to Eq.~(\ref{sab}) positive (negative) values of
$S_{ab}(k)$ indicate attraction (repulsion) in momentum space,
see also Ref.~\cite{fehske-ktnp05}.

Starting at low densities ($r_s = 10$), the negative values of the
e-e and h-h structure factors
at small momenta (large distances) originate from
the strong Coulomb repulsion of quasi-free equally charged particles.
The maximum of $S_{eh}$, as well as the modulations of $S_{ee}$ and $S_{hh}$
at finite $k$, are indicative of exciton formation.
Excitons set a new length scale in the structure factor at about
 0.2 [$1/2a_B$]. As expected,
these signatures are washed out at higher temperatures, where excitons
break up. 
At high density ($r_s = 1$) the large mass ratio between
electrons and holes is responsible for the different
behavior of $g_{ee}$ and $g_{hh}$. 
This is especially true for low temperatures when the
electrons are perfectly delocalized but the holes still have
fluid-like short-range correlations as noted above. 
The value of the normalization constant (denominator)
in Eq.~(\ref{sab}) is responsible for the different magnitude of
the $S_{hh}$ at $T = 50 K$ and $T = 200 K$.

The e-e- and h-h PDF and structure factors 
presented in Figs.~\ref{geehh} and \ref{seehh1}
were averaged over the spin degree of freedom of the
particles. Our DPIMC simulations, however, allow direct inspection of the
spin effects as well. Especially for the case of large degeneracy
($n \lambda^3\gg 1$) we expect qualitatively different behavior of
$g^{\uparrow \uparrow}_{aa}$ and $g^{\uparrow \downarrow}_{aa}$ at
small distances, due to the influence of the Fermi statistics ($a=e,h$). This 
is confirmed by Fig.~\ref{gseehh}. In other words, we clearly see the
exchange-hole known from the ideal Fermi gas, but here it appears in 
a strongly interacting e-h plasma. In contrast,
for the much heavier holes the quantum statistical repulsion
is much less pronounced, and the decay of $g_{hh}$ for $r\to 0$ is mainly triggered by the
(spin-independent) Coulomb interaction.

The snapshots depicted in Fig.~\ref{Snpsht} have shown that in the
density regime $2 \leq r_s \leq 6$ the plasma contains a large fraction of
bi-excitons. Now, Fig.~\ref{gseehh} confirms this conclusion (consider, e.g.,
the case $r_s=4$): There is not much difference in $g^{\uparrow \uparrow}_{hh}$ and $g^{\uparrow \downarrow}_{hh}$ - both functions exhibit a similar peak at a hole-hole distance of
about $3a_B$. In contrast, the electron behavior is strongly spin dependent:
There is a high probability to encounter two electrons with different spin
projections between two holes 
(at a distance $1a_B\dots 2a_B$) -- 
which is the expected behavior known 
from bi-excitons (or the hydrogen molecule). 

The corresponding charge structure factors clearly show
that the difference between electrons and holes with respect to the
spin correlations are most pronounced in the regime where bound states
are formed but, of course, the differences are less important at very small
and large $k$ (corresponding to large and small distances, respectively).
We found attraction (repulsion)
for opposite (equal) spin projections for both particles as $k\to 0$,
except for the case of high densities,
where the hole fluid forms. Here, in the vicinity of
the Mott transition, the difference
between the structure factor for electrons and holes with same
and opposite spin projections is notably smaller; they are close 
to their spin-averaged values given in Fig.~\ref{seehh1} for 
$T = 50$~K and $r_s = 2$.

\subsubsection{Degree of ionization. Mott density}

The above analysis of the snapshots and PDFs has indicated that, 
in a broad range of densities, 
the e-h-plasma is partially ionized containing a varying fraction of free 
and bound electrons and holes. 
The value of the critical Mott density (corresponding 
to the Brueckner parameter $r_s^{\rm Mott}$) where 
excitons vanish due to pressure ionization 
is essential for the occurrence of hole crystallization.
In particular, the critical mass ratio was found to directly 
depend on $r_s^{\rm Mott}$ as 
 $M^{cr}+1=r_s^{cr}/r_{se}^{\rm Mott}$, where $r_{se}\equiv 2^{1/3}r_s$ and 
$r_s^{cr}\approx 100$ is the 
critical value for Coulomb crystallization in a one-component 
plasma~\cite{rsc_ocp}. The value of $r_s^{\rm Mott}$
depends in a complicated way on the 
structure of the excitons (which may be quite different for 
different materials) and on many-body effects such as screening and 
bound-state renormalization.
This is a very complex problem which has been discussed 
in a variety of approximations the accuracy of which, 
however, is difficult to assess (see 
e.g. Ref.~\cite{green-book} for an overview). 
Based on our DPIMC simulations avoiding 
any simplifying assumptions, we have the 
possibility to obtain a consistent 
many-body result for $r_s^{\rm Mott}$.  
To this end, we first need to find a definition for the degree of 
ionization which is applicable to our simulations.

The main problem of PIMC simulations in configuration space is 
that there is, in principle, 
no clear subdivision into bound and free ``components'' possible. 
Electron-hole correlations
arising from bound and scattering states contribute to the same quantities, 
such as energy, 
equation of state or PDF. Note that the usual subdivision in 
energy space into bound and 
scattering states according to negative and positive relative 
pair energies $E_{\alpha}$ 
also breaks down in the case of strong correlations: 
Eigenvalues $E_{\alpha}$ and wavefunctions $\Psi_{\alpha}(r)$ 
are being renormalized, 
$\Psi_{\alpha} \rightarrow {\tilde \Psi}_{\alpha}; 
E_{\alpha} \rightarrow {\tilde E}_{\alpha}$. In the vicinity of the Mott point, the distinction between bound and scattering states becomes meaningless, 
bound state levels merge into the scattering continuum \cite{green-book}.

Nevertheless, a rough estimation
of the fraction $1-\alpha^{ion}$ of e-h bound states ($\alpha^{ion}$ denotes the degree of ionization) can be obtained
by analyzing the PDF $g_{eh}$. In particular, 
at low temperatures $k_BT \ll E_B$ and far below 
the Mott density, the plasma will consist of 
excitons in the ground state $E_1$,  i.e.
$g_{eh}(r)\sim |\Psi_{1}(r)|^{2}$ (see~\cite{filinov-etal.00pla}). 
In the general case, the PDF will contain a superposition (mixed state) of all renormalized eigenfunctions, weighted with the Boltzmann factor,
\begin{equation}
g_{eh}(r)\sim \sum_{\alpha}|{\tilde \Psi}_{\alpha}(r)|^{2}\exp(-\beta
{\tilde E}_{\alpha}).
\label{geh_ren}
\end{equation}
While the scattering states are delocalized, the bound state wavefunctions are localized 
in space leading to an increase of $g_{eh}$ beyond 
the result for an ideal plasma $g^{id}_{eh} \equiv 1$. 
Due to the normalization of $g_{eh}$ values of $g_{eh}>1$ 
at small e-h distances must be compensated by a depletion, 
$g_{eh} < 1$, at larger distances. Now, recall that $g_{eh}$ is related to the probability 
density by $P_{eh}(r) \sim r^2g_{eh}(r)$ which, 
in the case of excitons, is strongly 
enhanced around $r_{eh}=1a_B$, cf. Fig.~\ref{geehh}. 
Hence we can use as a ``pragmatic'' 
definition of the fraction of bound states (this idea is due to N.~Bjerrum who used it 
very successfully in the theory of electrolytes \cite{bjerrum})
the probability of e-h-pairs being at 
small distances (where $g_{eh}>1$):
\begin{eqnarray}
\frac{N_{eh}^b}{N_{eh}^b+N_{eh}^c} \equiv 1-\alpha^{ion} \approx
\frac{\int_{0}^{r^b}r^2\,[g_{eh}(r)-1]\,dr}{\int_{0}^{r^b}
r^2\,g_{eh}(r)\,dr}\,.
 \label{alfareh}
\end{eqnarray}
Here $r_b$ is the second 
zero of $r^2(g_{eh}-1)$ which is located to the right of the 
exciton peak. By subtracting $\int_{0}^{r^b}r^2\,dr $ in the 
nominator, the uncorrelated contributions are 
eliminated from the full probability density.
The denominator accounts for the normalization 
(giving the full probability of particles being in bound or scattering states).
Of course, this definition is only qualitative since it does not exclude attractive scattering states and thus may slightly overestimate the 
``true'' bound state fraction.

An analogous ratio can be defined for the hole-hole
bound-state (bi-exciton) fraction
\begin{eqnarray}
\frac{N_{hh}^b}{N_{hh}^b+N_{hh}^c} = \frac{\int_{r^{b'}}^{r^b}r^2\,[
g_{hh}(r)-1]\,dr}{\int_{r^{b'}}^{r_b}r^2\, g_{hh}(r)\,dr},
 \label{alfarhh}
\end{eqnarray}
with $[r^{b'},r_b]$ being the interval where $r^2\,( g_{hh}(r)-1)$ is
positive.

We expect that the definition (\ref{alfareh}) is well suited for a numerical determination 
of the Mott density as the density, where the plasma becomes dominated by free particle 
behavior. We will identify $r_s^{\rm Mott}$ as the density where the bound state fraction 
falls below $5\dots 10\%$, i.e. $\alpha^{ion}=90\dots 95\%$.

\begin{figure}[t]
\includegraphics[width=7cm,angle=0,clip=true]{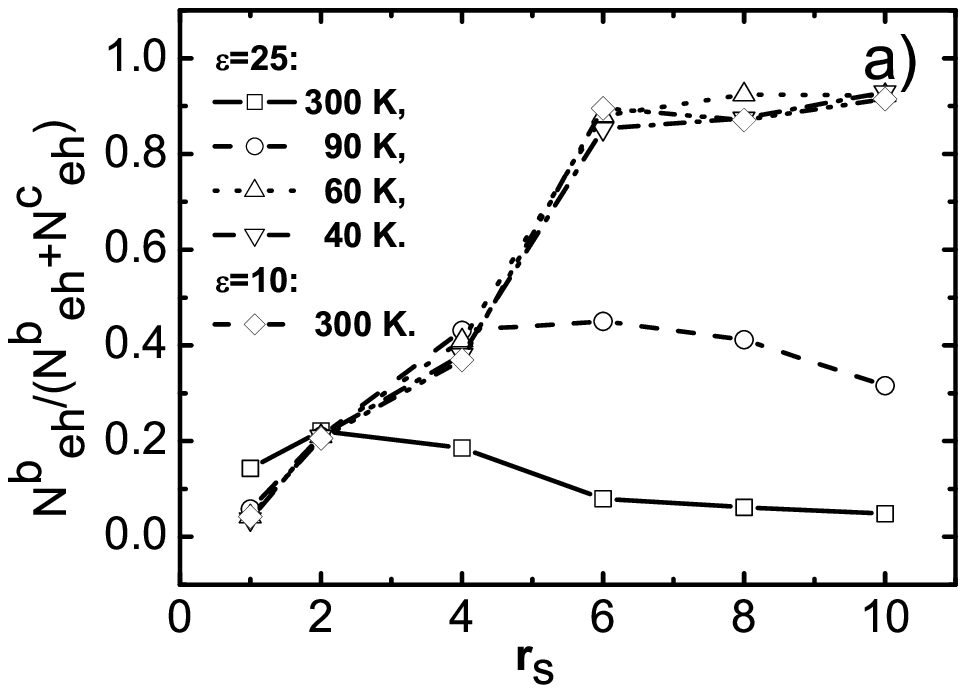}
\includegraphics[width=7cm,angle=0]{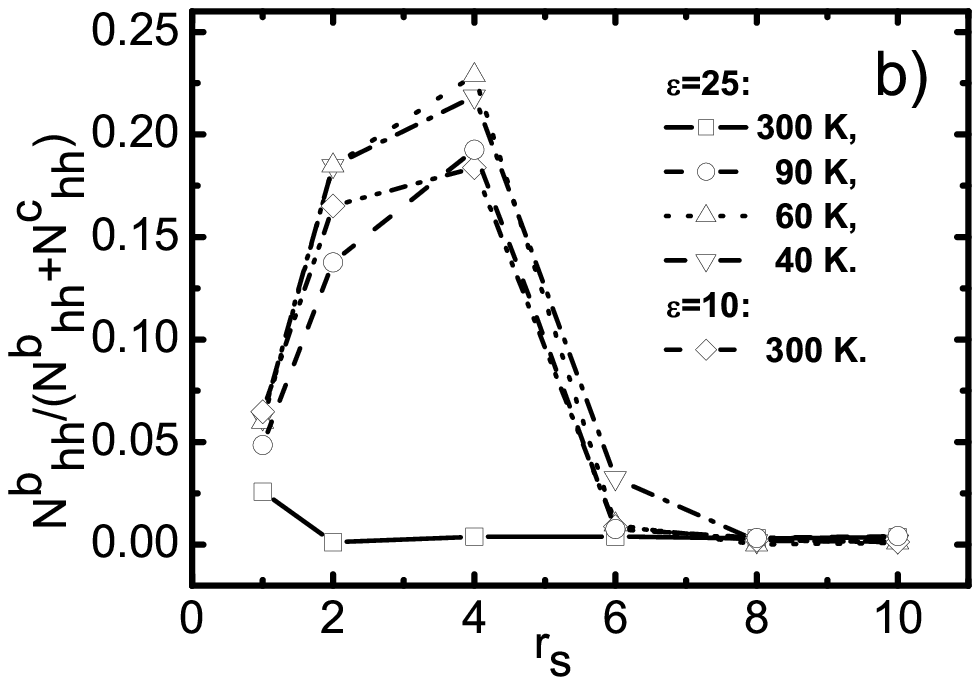}
\includegraphics[width=7cm,angle=0,clip=true]{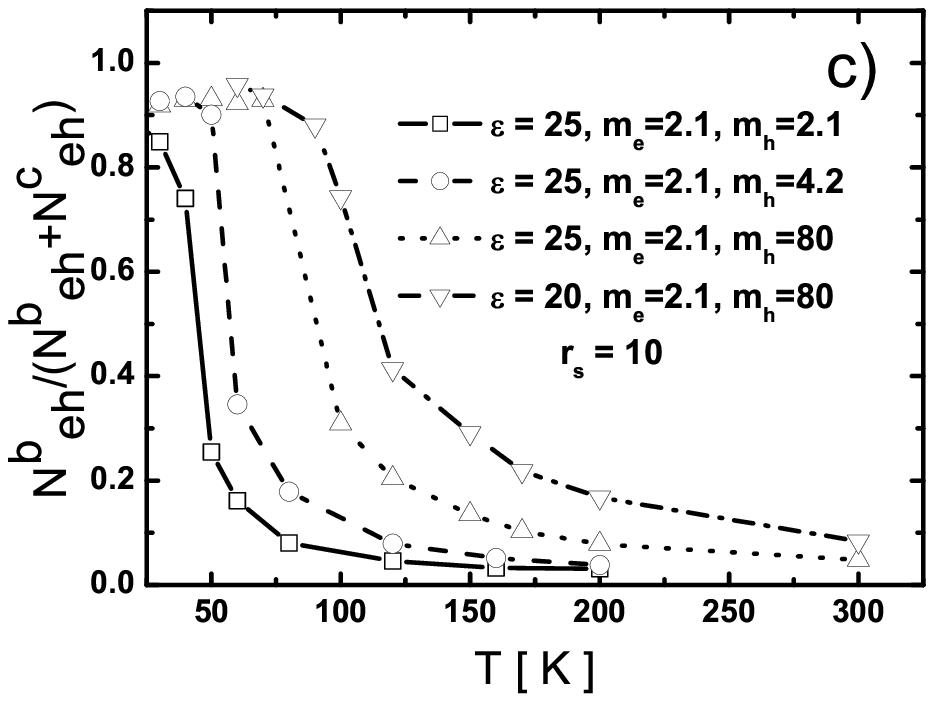}
\caption{Fraction of the electron-hole (top left, a)) and hole-hole
bound states (top right, b)) versus Brueckner parameter
$r_s$ for different temperatures and dielectric constants. 
Bottom Fig. (c) illustrates the temperature induced break up
of excitons for different mass ratios and dielectric constants.}
\label{cmpexp}
\end{figure}

Figure~\ref{cmpexp} presents our DPIMC results for the e-h bound state fraction according to
Eqs.~(\ref{alfareh}), (\ref{alfarhh}). At low densities ($r_s>6$)
and low temperatures ($T<100$~K) practically all electrons and
holes are bound in excitons, i.e. bi-excitons and many-particle
clusters are absent. Increase of temperature, $T\geq 100 K$,
leads to a strong ionization of excitons, i.e. free particles
dominate. In the intermediate density regime ($2<r_s<6$) 
and for low temperatures ($T<100$~K), 
the largest fraction (up to about $20 \%$) of h-h bound states
(bi-excitons and many-particle clusters) is observed. It is 
interesting to note that at higher temperatures ($T\sim 100$K$\dots
300 $~K) the exciton fraction increases a bit for higher density.

A decrease of the bound state fraction below $5\%$ is observed for $r_s\approx 1$ 
which we, therefore, identify with the Mott density, $r_s^{\rm Mott}\approx 1$ with 
an error of about $30\%$ (which corresponds to $r_{se}\approx 1.2$ which was used 
 in Ref. \cite{Bonitz_PRL05}). 
From this result, we confirm the value of the critical mass 
ratio (see above) $M^{cr}\approx 83$ which is in the range of previous predictions, e.g. 
\cite{mcr_data}. Since also the value for $r_s^{cr}$ is expected to have an 
error of about $20\%$, a total uncertainty of about $50\%$ has to be 
expected. This means that hole crystallization might occur already below 
$M\approx 80$ which underlines again the interest 
in the ${\rm Tm[Se_xTe_{1-x}]}$ system.

Let us, therefore, come back to the experimental phase diagram, 
Fig.~\ref{newfig5}. There the excitonic-rich phase should exist  
up to a temperature of about $250$K. Then it is of course  
interesting to calculate the temperature 
dependence of the e-h bound state fraction in this regime
(cf. the results for $r_s=10$, bottom part Fig.~\ref{cmpexp}).
Although our simulations do not exhibit a sharp 
transition from an excitonic to an ionized ``phase'', we 
can again formally use a bound state fraction of 10~\% as 
the boundary of the exciton dominated phase. Then,
at low density, e.g. for $r_s=10$, these bound states
will be stable up to $T=150\dots 250$~K. Note that these 
values are quite sensitive 
to $\varepsilon$ which is not precisely known. 
We therefore have performed simulations 
for two slightly different values. 
Increase of $\varepsilon$ leads to a more weakly bound
excitons and consequently reduces the ionization temperature.
The same tendency is observed if the mass ratio $M$ is lowered.
Fig.~\ref{cmpexp} shows that the ratio of temperatures
corresponding to $\alpha^{ion}=50~\%$ is
the same as the corresponding ratio of the reduced
masses $m_r=m_e m_h/(m_e+m_h)$ (keeping $\varepsilon$ fixed).
For example, a ratio of about 2 is observed for the
data belonging to $m_h=80$ and $m_h=2.1$, and 1.5
for those belonging to $m_h=80$ and $m_h=4.2$.
Clearly, as expected for an isolated exciton, 
the binding energy is proportional
to $m_r$. Note however that in the present case the excitons
are embedded into an e-h plasma which influences the bound state spectrum,
so the recovery of the single exciton behavior is not a trivial result.

We can also give a rough estimate of the critical electron density where the 
transition from an insulator (built up of
e-h bound states) to a (semi-) metal takes place,
again using a value of $\alpha^{ion}=90\%$ 
as a criterion. 
The critical value of $r_s \sim 1$  corresponds to a particle density of the
order of $10^{21}$~cm$^{-3}$ which is in reasonable agreement
with the estimates from the experiments. 

\subsection{Increased hole localization with increasing mass ratio}

After we have confirmed the pressure induced break up of excitons and determined the 
Mott parameter $r_s^{\rm Mott}$ in ${\rm Tm[Se_xTe_{1-x}]}$ with 
a mass ratio of $M\approx 40$, we can now return to the question of strong h-h correlations 
at high density and the possibility of hole crystallization.
To this end we have performed DPIMC simulations 
at low temperatures and high densities, well above the Mott point,
and varied the mass ratio over two orders of magnitude.

\begin{figure}[t]
\includegraphics[width=10cm,angle=0,clip=true]{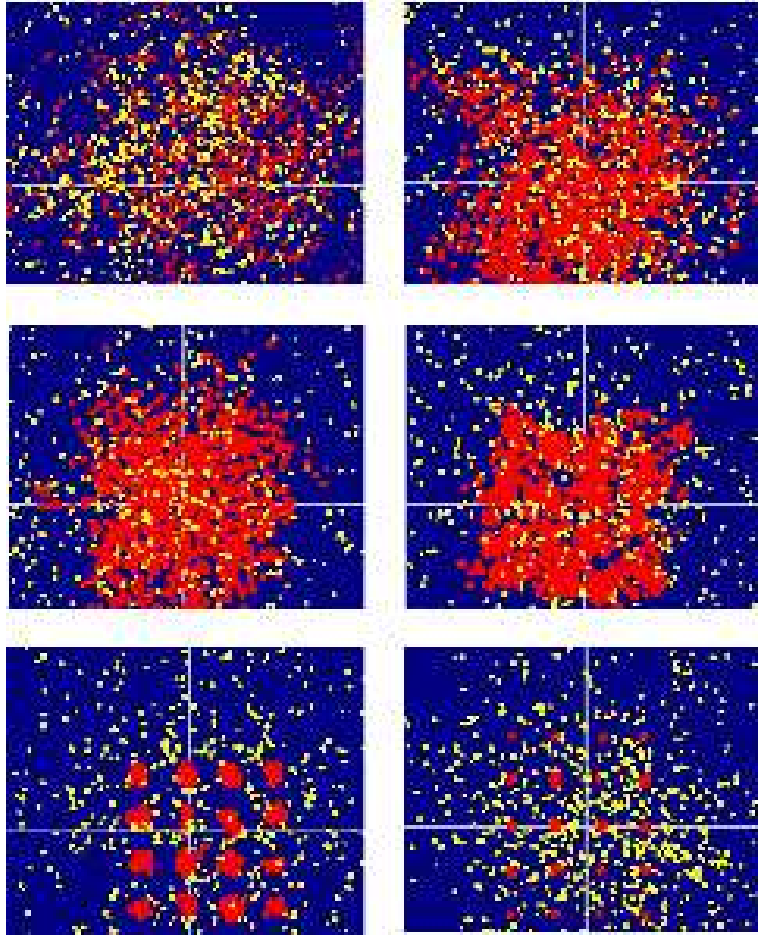}
\caption{(Color) Snapshots of the electron-hole plasma at high density 
($r_s=0.5$) and $T/E_B=0.064$. The pictures 
show the spin-averaged electron (yellow) 
and hole (red) beads configuration for different values 
of the mass ratio $M$ 
(from top left to bottom right): 5, 12, 25, 50, 100, 400.}
\label{snap_m}
\end{figure}

Fig.~\ref{snap_m} shows the results for $r_s=0.5$ and $k_BT/E_B=0.064$ 
with $M$ ranging from $5$ to $400$. To simplify the analysis we 
do not mark electrons and holes having 
different spin projections with different colors 
(we found that, for hole crystallization, 
spin effects are of minor importance). Since the density is 
an order of magnitude higher than the Mott density, electrons and 
holes are in the plasma state. For all values of $M$ the electrons are 
completely delocalized, forming a Fermi gas which is very weakly correlated. 
At the same time, the nature of the hole state changes 
drastically increasing of $M$. Initially ($M=5$), the holes 
are also in a Fermi gas-like state where individual 
holes penetrate each other. An increase of $M$ up to $50$ 
leads to continuous growing hole localization. At $M=50$ individual holes 
can be distinguished: Their DeBroglie wavelength $\lambda_h$ 
becomes smaller than the average distance between two holes $d_h$. A further 
increase of $M$ to 100 leads to an additional reduction 
of $\lambda_h$ by a factor $\sqrt{2}$ (with $d_h$ unchanged) and, 
consequently, to hole localization. The holes form a 
regular lattice - a Coulomb crystal (bottom left Fig.). 
Increasing the mass ratio further, the lattice becomes 
more rigid (bottom right Fig.), until the holes become practically point-like  
(cf. Fig.~\ref{newfig4}). Based on these simulation results 
we can conclude that, at these values of density and temperature, hole crystallization occurs between $M=50$ and $M=100$, which 
is in surprisingly good agreement with the analytical 
estimate for $M^{cr}$ based on the 
Mott density obtained above for the same parameters.

Obviously, the snapshots allow for a very 
rough determination of the critical mass 
ratio only. A more accurate estimate  
can be obtained by analyzing the relative distance fluctuation of  
the holes in dependence on $M$. These results, 
reported in Ref.~\cite{bonitz-etal.06jpa}, 
confirm the above result for $M^{cr}$. We  
also  have studied the thermal properties of 
the hole crystal which arise to a temperature 
dependence of $r_s^{\rm Mott}$. As seen 
from Fig.~\ref{cmpexp} (left part)  
the bound state fraction isotherms  
has reached saturation at about $60$K 
(corresponding to slightly more than $0.1 E_B$). Thus our
result for $M^{cr}$ will not change significantly at lower temperatures,
thus reflecting the ground state 
behavior. The full phase diagram of the hole crystal was published in 
Ref.~\cite{Bonitz_PRL05}.

\section{Discussion}\label{summary}
In this paper we have presented a numerical (computer simulation) analysis of strong Coulomb correlations in dense three-dimensional two-component plasmas at rather low temperatures. We were in particular interested in systems with a mass ratio $M\sim 40$ which is intermediate between ``usual'' semiconductors ($M\sim 2\dots 10$) and ``conventional'' plasmas (such as hydrogen with 
$M \ge 1836$). Two-component Coulomb systems with this mass ratio behave in many aspects like plasmas, with the main difference that quantum and spin properties of the heavy particles (holes) cannot be neglected. Such systems might be realized in intermediate valence semiconductors under pressure. 

From a plasma physics standpoint these are very interesting materials because they allow 
to investigate nontrivial high density phenomena of current interest, such as pressure ionization (Mott effect), plasma phase transition and 
metal-insulator transition. 
These effects exist in ``conventional'' plasmas only above 
a density of the order of $1$g cm$^{-3}$ which, meanwhile, can be achieved in laser or ion 
beam compression experiments, however, only for short periods of time which makes precision 
measurements of the plasma properties very difficult. The same physical effects can be 
observed in the above mentioned semiconductor materials under stationary 
(equilibrium) conditions at densities 
and temperatures which are easily accessible in experiments. In fact, the (qualitative) phase diagrams of dense plasmas and electron-hole plasmas are readily translated from one to another by rescaling the binding energy and the Bohr radius \cite{bonitz_pj02,bonitz-etal.03jpa}. For this reason, the phase diagram of Tm[Se,Te], shown in Fig.~\ref{newfig5}, is of interest also for 
the physics of dense partially ionized plasmas. 

In this paper, we concentrated on the central part of this 
phase diagram, where a large fraction of bound states exists, on the Mott transition  and on the 
high density effects (hole crystallization).
A theoretical treatment of these effects is very difficult, because many-body effects such as bound state formation, screening and quantum effects have to be taken into account self-consistently. 
While for such complex systems analytical methods fail, our first principle path integral Monte Carlo approach is well suited. In order to avoid additional approximations which are particularly questionable in the region of the Mott point, we have used direct fermionic simulations. With an 
improved treatment of the spin statistics we were able to present reliable simulations. Restricting 
the simulations to temperatures above $6\%$ of the exciton binding energy and densities to 
 values not significantly higher than the Mott density allowed us to avoid serious difficulties 
arising from the fermion sign problem.

Most notably, we have shown that, above the Mott point, two-component plasmas with large mass anisotropy show interesting Coulomb correlation phenomena: with increasing density holes can 
undergo a 
phase transition to a Coulomb liquid and to a Wigner crystal which are embedded into a degenerate 
electron Fermi gas. Such crystals are expected to exist in White Dwarf stars where the mass ratio exceeds $10^{4}$. However, crystal formation in a two-component plasma should be possible also for the light elements, such as hydrogen \cite{filinov-etal.00jetpl,militzer06} and helium and, more generally, for plasmas with mass ratios as low as 80 
(cf.~\cite{Bonitz_PRL05}). This should be possible to achieve in the semiconductor materials announced in this paper. 
More subtle questions, such as the symmetry of the crystal and its energy
cannot yet been answered conclusively, mainly 
because of the yet too small size of the simulations 
(50 electrons and 50 holes are presently feasible). Therefore, in order to
obtain more accurate data, e.g. for the internal energy of a
macroscopic two-component plasma at very high density, a
significant increase of the simulation size would be highly desirable.

\section*{Acknowledgements}
We thank A.~Abrikosov and H.~E.~DeWitt for stimulating 
discussions on hole crystallization. We are 
grateful to P.~Wachter and B.~Bucher for valuable conversations and details 
regarding his measurements on Tm[Se,Te] as well as for the permission to use Fig.~\ref{newfig5}. 
Computations
were performed on the compute-clusters of the Universities
Greifswald and Kiel. This work has been supported by the
Deutsche Forschungsgemeinschaft through SFB TR-24 and SFB 652 and
by RF President Grants No. MK-3993.2005.8 and the RAS program No. 17
and in part by Award No. Y2-P-11-02 of the U.S.
Civilian Research $\&$ Development Foundation for the
Independent States of the Former Soviet Union (CRDF) and
of Ministry of Education and Science of Russian
Federation, and RF President Grant NS-3683.2006.2 for
governmental support of leading scientific schools.

\begin{appendix}
\section{Path integral representation of thermodynamic quantities}
\label{theory_a}
Let us consider a neutral two-component plasma consisting of $N_e=N_h=N$ electrons and holes in equilibrium with the hamiltonian,
${\hat H}={\hat K}+{\hat U}^c$,
containing kinetic energy ${\hat K}$ and Coulomb interaction energy
${\hat U}^c = {\hat U}_{hh}^c + {\hat  U}^c_{ee} + {\hat
U}^c_{eh}$ contributions. The thermodynamic properties in the
canonical ensemble with given temperature $T$ and fixed volume $V$ are fully described by the density operator ${\hat \rho} = e^{-\beta {\hat H}}/Z$ with the partition function~(\ref{q-def}). 

Pressure and internal energy follow from
\begin{eqnarray}
\beta p &=& \frac{\partial {\rm ln} Z}{\partial V} = \left[\frac{\alpha}{3V}
\frac{\partial{\rm ln} Z}{\partial \alpha}\right]_{\alpha=1}, \label{p_gen}
\\
\beta E &=& -\beta \frac{\partial {\rm ln} Z}{\partial \beta},
\label{e_gen}
\end{eqnarray}
where $\alpha= L/L_0$ is a length scaling parameter.

The density matrix of interacting quantum
systems is can be constructed using a path integral
approach~\cite{feynman-hibbs} based on the operator identity
$e^{-\beta {\hat H}}= e^{-\Delta \beta {\hat H}}\cdot
e^{-\Delta \beta {\hat H}}\dots  e^{-\Delta \beta {\hat H}}$,
where $\Delta \beta = \beta/(n+1)$, which allows us to
rewrite the integral in Eq.~(\ref{q-def})
\begin{eqnarray}
&&\sum_{\sigma} \int\limits dq^{(0)}\,
\rho(q^{(0)},\sigma;\beta) =
\int\limits  dq^{(0)} \dots
dq^{(n)} \, \rho^{(1)}\cdot\rho^{(2)} \, \dots \rho^{(n)} \times\nonumber\\
&&\qquad\sum_{\sigma}\sum_{P_e} \sum_{P_h}(\pm 1)^{\kappa_{P_e}+ \kappa_{P_h}}
{\cal S}(\sigma, {\hat P_e}{\hat P_h} \sigma_{a}^\prime)
{\hat P_e} {\hat P_h}\rho^{(n+1)}\big|_{q^{(n+1)}= q^{(0)}, \sigma'=\sigma}\,.
 \label{rho-pimc_a}
\end{eqnarray}
The spin gives rise to the spin part of the density matrix (${\cal
S}$) with exchange effects accounted for by the permutation
operators  $\hat P_e$ and $\hat P_h$ acting on the electron and
hole coordinates $q^{(n+1)}$ and spin projections $\sigma'$. The
sum is over all permutations with parity $\kappa_{P_e}$ and
$\kappa_{P_h}$. In Eq.~(\ref{rho-pimc_a}) the index $k=1\dots n+1$
labels the off-diagonal high-temperature density matrices
$\rho^{(k)}\equiv \rho\left(q^{(k-1)},q^{(k)};\Delta\beta\right) =
\langle q^{(k-1)}|e^{-\Delta \beta {\hat H}}|q^{(k)}\rangle$.
Accordingly each particle is represented by a set of $n+1$ coordinates 
(``beads''), i.e. the
whole configuration of the particles is represented by a
$3(N_e+N_h)(n+1)$-dimensional vector
$\tilde{q}\equiv\{q_{1,e}^{(0)}, \dots q_{1,e}^{(n+1)},
q_{2,e}^{(0)}\ldots q_{2,e}^{(n+1)}, \ldots q_{N_e,e}^{(n+1)};
q_{1,h}^{(0)}\ldots q_{N_h,h}^{(n+1)} \}$ (see Fig.~\ref{beads}).

To determine the energy  in the path integral
representation~(\ref{rho-pimc_a}) each high-temperature density matrix has to be
differentiated in turn (here we extend our earlier hydrogen results of Refs.~\cite{FiBoEbFo01,zamalin}):\\
\begin{eqnarray}
\hspace*{-1cm}\beta E &=& - \frac{1}{Z} \int\limits dq^{(0)} \dots
dq^{(n)}\,\sum_{\sigma}\sum_{P_e} \sum_{P_h} (\pm 1)^{\kappa_{P_e}+
\kappa_{P_h}} \,{\cal S}(\sigma, {\hat P_e} {\hat P_h} \sigma')\,
\nonumber\\
&&\;\times  \sum_{k=1}^{n+1}\rho^{(1)}
\dots
\rho^{(k-1)}
\left[\beta\frac{\partial \rho^{(k)}}{\partial \beta}\right]
\rho^{(k+1)} \, 
\dots \, \rho^{(n)}{\hat P_e} {\hat P_h} \rho^{(n+1)}
\bigg|_{q^{(n+1)}=q^{(0)},\, \sigma'=\sigma}\,. \label{e-pimc_a}
\end{eqnarray}
One can show that the matrix elements $\rho^{(k)}$ can be rewritten as
\begin{eqnarray}
\hspace*{-1cm}\rho^{(k)}&\equiv&\langle q^{(k-1)}|e^{-\Delta \beta {\hat
H}}|q^{(k)}\rangle =
\int d{\tilde p}^{(k)}d{\bar p}^{(k)}\, \langle
q^{(k-1)}|e^{-\Delta \beta {\hat U}^c}|{\tilde p}^{(k)}\rangle
\langle {\tilde p}^{(k)}|e^{-\Delta \beta {\hat K}} |{\bar
p}^{(k)}\rangle
\, \nonumber\\
&&\hspace*{6cm}\times \langle {\bar p}^{(k)}| \,e^{-\frac{\Delta
\beta^2}{2}[{\hat K},{\hat U}^c]} \, \dots|q^{(k)}\rangle\,,
\label{rho_ku_a}
\end{eqnarray}
where ${\tilde p}^{(k)}({\bar p}^{(k)})$ are conjugate variables to
$q^{(k-1)}(q^{(k)})$. Evaluating the derivatives in Eq.~(\ref{e-pimc_a}) further, it is convenient to introduce dimensionless integration variables
$\eta^{(k)}=(\eta_h^{(k)},\eta_e^{(k)})$, where
$\eta_a^{(k)}=\kappa_a(q^{(k)}_a-q^{(k-1)}_a)$ for $k=1,\dots, n$,
and $\kappa_a^2\equiv m_a /(2\pi \hbar^2 \Delta  \beta
)=1/\lambda_{\Delta ,a}^2$. The main
advantage is that differentiation of the density matrix now affects only
the interaction terms
\begin{eqnarray}
\beta \frac{\partial \rho^{(k)}}{\partial \beta} &=& -\beta
\frac{\partial [\Delta\beta \cdot U^c(X^{(k-1)})]}{\partial \beta}
\rho^{(k)} +\beta {\tilde \rho}_{\beta}^{(k)}, \label{rho-prime_a}
\end{eqnarray}
where
\begin{equation}
\hspace*{-1cm}{\tilde \rho}_{\beta}^{(k)} = \int d{p}^{(k)}\, \langle
X^{(k-1)}|e^{-\Delta \beta {\hat U}^c}|{p}^{(k)}\rangle
e^{-\frac{\langle {p}^{(k)}|{p}^{(k)}\rangle}{4\pi(n+1)}}
\,
\langle {p}^{(k)}|\frac{\partial}{\partial \beta}
\,e^{-\frac{(\Delta\beta)^2}{2}[{\hat K},{\hat U}^c]} \,
\dots|X^{(k)}\rangle  \label{rho-tilde_a}
\end{equation}
with $p_a^{(k)}={\tilde p}_a^{(k)}/(\kappa_a\hbar)$,
$p{(k)}\equiv (p_h^{(k)},p_e^{(k)})$, and use has been made of Eq.
(\ref{rho_ku_a}).
Furthermore
$X^{(0)}\equiv (\kappa_h q_h^{(0)},\kappa_e q_e^{(0)})$,
$X^{(k)}\equiv (X_h^{(k)},X_e^{(k)})$ with $X_a^{(k)}=\kappa_a
q_a^{(0)}+\sum_{l=1}^k \eta^{(l)}_a$ ($k$ runs from $1$ to
$n$), and $X^{(n+1)}\equiv (\kappa_h q_h^{(n+1)},\kappa_e
q_e^{(n+1)})=X^{(0)}$.

For $k=n+1$, we have
\begin{equation}
\beta \frac{\partial}{\partial \beta}
\, {\hat P_e} {\hat P_h} \rho^{(n+1)} =
-\beta \frac{\partial \Delta\beta \cdot U^c(X^{(n)})}{\partial
\beta} {\hat P_e} {\hat P_h}\rho^{(n+1)} + \beta {\hat P_e} {\hat P_h}
{\tilde \rho}_{\beta}^{(n+1)}\,.
\label{rho-prime1}
\end{equation}
Then, together with Eq.~(\ref{e-pimc_a}), we obtain for the energy
\begin{eqnarray}
\beta E &=& \frac{3}{2}(N_e+N_h) -
 \frac{1}{Z}
\frac{1}{\,\lambda_h^{3N_h}\lambda_e^{3N_e}} \int\limits_{V}
dq^{(0)} d\eta^{(1)} \dots d\eta^{(n)} \,
\nonumber\\
&\times & \sum_{\sigma}\sum_{P_e} \sum_{P_h} (\pm
1)^{\kappa_{P_e}+ \kappa_{P_h}} \,{\cal S}(\sigma, {\hat P_e}
{\hat P_h} \sigma')
\nonumber\\
&\times & \Bigg\{\sum_{k=1}^{n+1}\rho^{(1)}\dots\rho^{(k-1)}
\left[ \beta {\tilde \rho}_{\beta}^{(k)} -\beta \frac{\partial
\Delta\beta \cdot
U^c(X^{(k-1)})}{\partial \beta}\rho^{(k)}
 \right]
\nonumber\\
&\times & \rho^{(k+1)} \, \dots \,\rho^{(n)} {\hat P_e} {\hat P_h}
\rho^{(n+1)} \Bigg\}\bigg|_{X^{(n+1)}=X^{(0)},\, \sigma'=\sigma}.
\label{pimc1}
\end{eqnarray}
This way the derivatives of the density matrix have been
calculated and we turn to the next point:
Finding approximations for the high-temperature density matrices $\rho^{(k)}$.

\section{High-temperature asymptotics for the density matrix. 
Kelbg potential}\label{theory_b}
In this section we discuss approximations for the high-temperature
density matrix that can be used for efficient DPIMC simulations.
This involves effective quantum pair potentials $\Phi_{ab}$, which are
approximated by the Kelbg potential (see also previous work~\cite{FiBoEbFo01}). 
\subsection{Pair approximation and Kelbg potential}
The N-particle high-temperature density matrix is expressed 
in terms of two-particle
density matrices (higher order terms become negligible at sufficiently high
temperature, i.e. for large number $n$ of time slices) given by
(\ref{rho_ab}). 
This results from factorization into kinetic and interaction parts,
$\rho_{ab}\approx\rho_0^K\rho^{U^c}_{ab}$, which is exact in the classical
case, i.e. at sufficiently high temperature. The error made at finite 
temperature vanishes with the number of time slices as 
$1/n^{2}$ (cf.~Ref.\cite{FiBoEbFo01}).
The off-diagonal density matrix element (\ref{rho_ab}) involves an effective
pair interaction which is expressed approximately via its diagonal elements,
$\Phi^{OD}_{ab}(q_{p,a},q'_{p,a},q_{t,b}, q'_{t,b};\beta)\approx
[\Phi_{ab}(q_{p,a}-q_{t,b}; \beta)+\Phi_{ab}(q'_{p,a}-q'_{t,b};\beta)]/2$,
for which we use the familiar Kelbg potential\cite{Ke63,kelbg} 
(\ref{kelbg-d}). Note that the Kelbg
potential is finite at zero distance which is a consequence of quantum
effects. The validity of this potential as well as of the diagonal 
approximation is restricted to temperatures substantially higher than
the exciton binding 
energy.~\cite{afilinov-etal.04pre,ebeling_sccs05} which puts 
another lower bound on the number of time slices $n$.
For a discussion of other effective potentials, we refer to 
Refs.~\cite{KTR94,FiBoEbFo01,afilinov-etal.04pre,ebeling_sccs05}.

Summarizing the above approximations, we can conclude that
with the approximations (\ref{rho_ab}), (\ref{kelbg-d}) each of the 
high-temperature factors on the r.h.s. of Eq. (\ref{rho-pimc_a}),
carries an error of the order $1/(n+1)^2$.
Within these approximations, we obtain the result
\begin{eqnarray}
\rho^{(k)}=\rho_0^{(k)}e^{-\Delta \beta
U(X^{(k-1)})}\delta(X^{(k-1)}-X^{(k)})+{\cal O}[(1/n+1)^2]\,,
\label{kel}
\end{eqnarray}
where $\rho_0^{(k)}$ is the kinetic density matrix, and $U$ denotes
the sum of all interaction energies, each consisting of the
respective sum of pair interactions given by Kelbg potentials,
$U(X^{(k)})=U_{hh}(X_h^{(k)})+U_{ee}(X_e^{(k)})+U_{eh}(X_h^{(k)},X_e^{(k)})$.

\subsection{Estimator for the total energy}
Let us now return to the computation of thermodynamic functions and derive the
final expressions which follow from Eq.~(\ref{kel}) and which will be used in the simulations. First, we note that in Eq.~(\ref{pimc1}), special care has to be taken in performing the derivatives of the Coulomb potentials with respect to $\beta$: Products $\beta\frac{\partial \Delta\beta \cdot U^c(X^{(k-1)})}{\partial \beta}$
have a singularity at zero inter-particle distance which is
integrable but leads to difficulties in the simulations.
To assure efficient simulations we transform
the e-e, h-h and e-h contributions in the following way:
\begin{eqnarray}
&&\hspace*{-2cm}
\langle X^{(k-1)}|e^{-\Delta \beta {\hat K}}
|X^{k}\rangle \left[-\beta\frac{\partial}{\partial\beta}
\left(\Delta \beta
U^c(X^{(k-1)})\right) \right]
\nonumber\\
&&\approx  \int_0^1  d\alpha \int d\tilde{X}^{(k-1)} \langle
X^{(k-1)}|e^{-\Delta \beta \alpha{\hat K}}
|\tilde{X}^{(k-1)}\rangle
\left[-\beta\frac{\partial}{\partial\beta} \left(\Delta \beta
U^c(\tilde{X}^{(k-1)})\right) \right]
\nonumber\\
 && \qquad \times \langle \tilde{X}^{(k-1)}|e^{-\Delta \beta
(1-\alpha){\hat K}} |X^{k}\rangle +{\cal O}\left(1/(n+1)^2\right)
\, \nonumber\\
&&\approx  \langle X^{(k-1)}|e^{-\Delta \beta {\hat K}}
|X^{k}\rangle \left[-\beta\frac{\partial}{\partial\beta}
\left(\Delta \beta U(X^{(k-1)})\right) \right] + {\cal
O}\left(1/(n+1)^2\right). \label{uep}
\end{eqnarray}
This means, within the standard error of our approximation, ${\cal
O}\left(n^{-2}\right)$, we have replaced the sum of the
Coulomb potentials $U^c$ by the corresponding sum of Kelbg
potentials $U$, which is much better suited for MC simulations.
This result coincides with expressions, which can be obtained
if we first choose an approximation for the high-temperature
density matrices $\rho^{(k)}$ using Kelbg potential and then
take the derivatives.

Thus, our final result for the energy is
\begin{eqnarray}
\beta E &=& \frac{3}{2}(N_e+N_h) + \frac{1}{Z}
\frac{1}{\,\lambda_h^{3N_h} 
\lambda_e^{3N_e}}
\nonumber\\
&&\times \sum_{s=0}^{N_e} \sum_{k=0}^{N_h} \int\limits_{V} dq^{(0)}
d\eta^{(1)} \dots d\eta^{(n)} \,
\rho_{sk}(q^{(0)}, \eta^{(1)} \dots \eta^{(n)}, \beta)
\nonumber\\
&&\times \Bigg\{ \sum_{l=0}^{n}\Bigg[
\sum_{p=1}^{N_h}\sum_{t=1}^{N_e} \Psi_{eh}(|x^l_{pt}|) +
\sum_{p<t}^{N_h} \Psi_{ee}(|r^l_{pt}|)
+\sum_{p<t}^{N_e} \Psi_{hh}(|q^l_{pt}|)
 \Bigg]
\nonumber\\
&&+ \sum_{l=1}^{n}\Bigg[
 \sum_{p=1}^{N_h}\sum_{t=1}^{N_e}
D(x^l_{pt}) \frac{\partial \Delta\beta\Phi_{eh}}{\partial
|x^l_{pt}|}
 + \sum_{p<t}^{N_e}
C(r^l_{pt}) \frac{\partial \Delta\beta\Phi_{ee}}{\partial
|r^l_{pt}|} + \sum_{p<t}^{N_e} C(q^l_{pt}) \frac{\partial
\Delta\beta\Phi_{hh}}{\partial |q^l_{pt}|}
 \Bigg]
\nonumber\\
&&- \frac{1}{{\rm det} ||\psi^{n,0}_{pt}||_{sk}}
\frac{\partial{\rm \,det} || \psi^{n,0}_{pt} ||_{sk}}{\partial
\beta} \Bigg\},
\label{energy}
\end{eqnarray}
with
$C(r^l_{pt}) = \frac{\langle
r^l_{pt}|y^l_{pt}\rangle}{2|r^l_{pt}|}, \quad C(q^l_{pt}) =
\frac{\langle q^l_{pt}|\tilde{y}^l_{pt}\rangle}{2|q^l_{pt}|},
\qquad D(x^l_{pt}) = \frac{\langle
x^l_{pt}|y^l_{p}-\tilde{y}^l_{t}\rangle}{2|x^l_{pt}|}$ and
$\Psi_{ab}(x)\equiv \Delta\beta\partial [\beta'\Phi_{ab}(x,\beta')]/\partial\beta'|_{\beta'=\Delta\beta}$. Further, $\langle \dots | \dots \rangle$
denotes the scalar product, and $q_{pt}$, $r_{pt}$ and $x_{pt}$
are differences of two coordinate vectors: $q_{pt}\equiv
q_{p,h}-q_{t,h}$, $r_{pt}\equiv q_{p,e}-q_{t,e}$, $x_{pt}\equiv
q_{p,e}-q_{t,h}$, $r^l_{pt}=r_{pt}+y_{pt}^l$,
$q^l_{pt}=q_{pt}+\tilde{y}_{pt}^l$, $x^l_{pt}\equiv
x_{pt}+y^l_p-\tilde{y}^l_t$, $y^l_{pt}\equiv y^l_{p}-y^l_{t}$,
 $\tilde{y}^l_{pt}\equiv \tilde{y}^l_{p}-\tilde{y}^l_{t}$,
 with $y^l_t=\Delta\lambda_e\sum_{k=1}^{l}\eta^{(k)}_t$
 and
 $\tilde{y}^l_p=\Delta\lambda_h\sum_{k=1}^{l}\tilde{\eta}^{(k)}_p$.

The density matrices $\rho_{sk}$ appearing in Eq.~(\ref{energy}) are given by
\begin{eqnarray}
\rho_{sk} = C^s_{N_e} C^k_{N_h} \, e^{-\beta U}
\prod\limits_{l=1}^n \prod\limits_{p=1}^{N_e}
\prod\limits_{t=1}^{N_h} \phi^l_{p} \,\tilde{\phi}^l_{t}\, {\rm det}
\,||\psi^{n,0}_{pt}||_{sk}, \label{rho_s}
\end{eqnarray}
\begin{equation}
\mbox{where}\quad
U= \frac{1}{n+1}\sum_{l=0}^{n} \left\{U_e(X_e^{(l)}, \Delta
\beta)+U_h(X_h^{(l)},\Delta\beta)+
U_{eh}(X_h^{(l)},X_e^{(l)},\Delta\beta)\right\},
\end{equation}
and  $\phi^l_{t}\equiv \exp[-\pi |\eta^{(l)}_t|^2]$,
$\tilde{\phi}^l_{p}\equiv \exp[-\pi |\tilde{\eta}^{(l)}_p|^2]$. Notice
 that the density matrix (\ref{rho_s}) does not contain
an explicit sum over the permutations and thus no sum of terms
with alternating sign. Instead, the whole exchange problem is
contained in the following determinant which is a product of exchange 
matrices of electrons (index $s$) and holes (index $k$) where
$s$ ($k$) denotes the number of electrons and holes having the same spin
projections (or more details, we refer to Ref.~\cite{filinov76}),
\begin{eqnarray}
||\psi^{n,0}_{pt}||_{sk} = ||e^{-\frac{\pi}{\Delta\lambda_e^2}
\left|(r_a-r_b)+ y_a^n\right|^2}||_s \times
||e^{-\frac{\pi}{\Delta\lambda_e^2} \left|(r_a-r_b)+
y_a^n\right|^2}||_{N_e-s}
\nonumber\\
\times ||e^{-\frac{\pi}{\Delta\lambda_h^2} \left|(q_a-q_b)+
\tilde{y}_a^n\right|^2}||_k \times ||e^{-\frac{\pi}{\Delta\lambda_h^2}
\left|(q_a-q_b)+ \tilde{y}_a^n\right|^2}||_{N_h-k}\,.  \label{psi}
\end{eqnarray}

\subsection{Estimator for the equation of state}
In similar way, expressions for all other thermodynamic functions can be derived.
Here, we only provide the corresponding result for the equation of state,
\begin{eqnarray}
\frac{\beta p V}{N_e+N_h} &=& 1 - \frac{1}{N_e+N_h}
\frac{(3Z)^{-1}}{\,\lambda_h^{3N_h}
\lambda_e^{3N_e}}\nonumber\\
&&\times \sum_{s=0}^{N_e} \sum_{k=0}^{N_h} \int dq^{(0)} d\eta^{(1)}
\dots d\eta^{(n)} \,
\rho_{sk}(q^{(0)}, \eta^{(1)} \dots \eta^{(n)}, \beta) \nonumber\\
\nonumber\\
&&\times \Bigg\{ \sum_{p=1}^{N_h}\sum_{t=1}^{N_e} |x_{pt}|
\frac{\partial \Delta\beta\Phi_{eh}}{\partial |x_{pt}|}+
\sum_{p<t}^{N_h} |q_{pt}| \frac{\partial
\Delta\beta\Phi_{hh}}{\partial |q_{pt}|}
+ \sum_{p<t}^{N_e} |r_{pt}| \frac{\partial
\Delta\beta\Phi_{ee}}{\partial |r_{pt}|}
\nonumber\\
&&+\sum_{l=1}^{n}\left[ \sum_{p=1}^{N_h}\sum_{t=1}^{N_e}B(x^l_{pt})
\frac{\partial \Delta\beta\Phi_{eh}}{\partial |x^l_{pt}|}+
\sum_{p<t}^{N_e} A(r^l_{pt}) \frac{\partial
\Delta\beta\Phi_{ee}}{\partial |r^l_{pt}|} +
\sum_{p<t}^{N_h} A(q^l_{pt}) \frac{\partial
\Delta\beta\Phi_{hh}}{\partial |q^l_{pt}|}
 \right]
\nonumber\\
&&-\frac{\alpha}{{\rm det} ||\psi^{n,0}_{pt}||_{sk}}
\frac{\partial{\rm \,det} || \psi^{n,0}_{pt} ||_{sk}}{\partial
\alpha} |_{\alpha=1} \Bigg\},
 \label{eos}
\end{eqnarray}
with $B(x_{pt}) = \frac{\langle
x^l_{pt}|x_{pt}\rangle}{|x^l_{pt}|}$,  $A(r^l_{pt}) =
\frac{\langle r^l_{pt}|r_{pt}\rangle}{|r^l_{pt}|}, \quad$ and $
A(q^l_{pt}) = \frac{\langle q^l_{pt}|q_{pt}\rangle}{|q^l_{pt}|}$.

Eqs.~(\ref{energy}) for the total energy and~(\ref{eos}) for the pressure
are readily understood: the first terms on the r.h.s. correspond to the 
classical ideal gas part. The ideal quantum contribution,
in excess of the classical one, plus all
correlation contributions are contained in the integral terms. The Coulomb
correlation contributions arise from the terms with the Kelbg potentials
$\Phi_{ab}$, whereas the exchange contributions arise from the derivatives
of the exchange matrix (last term).

While there exist many alternative representations of the thermodynamic
functions, the main advantage of the present expressions 
Eqs.~(\ref{energy}) and~(\ref{eos}) for energy and pressure
is that the explicit sum over permutations has been converted into the 
spin determinant which can be computed very efficiently using standard
linear algebra methods. Furthermore, each of the sums in curly brackets in
Eqs.~(\ref{energy}) and~(\ref{eos}) is bounded when the number of
high-temperature factors increases ($n\rightarrow \infty$). Note
that Eqs.~(\ref{energy}) and~(\ref{eos}) contain the important
limit of an ideal quantum plasma in a natural way \cite{FiBoEbFo01,hermann}.

\end{appendix}

\end{document}